\documentclass[10pt, doublecolumn]{IEEEtran}  % Comment this line out if you need a4paper

\usepackage{tikz}
\usetikzlibrary{shapes,snakes}
\usepackage{empheq}
\usepackage{graphicx}
\usepackage{epstopdf}
\usepackage{multirow}
\usepackage{tabu}
\usepackage{array}
\usepackage{url}
\usepackage{amsmath}
\usepackage{amssymb}
\usepackage{amsthm}
\usepackage{amsfonts}
\usepackage{tikz}
\usepackage{bm}
\usepackage{algorithmic}
\usepackage[ruled,vlined,linesnumbered]{algorithm2e}
\usetikzlibrary{arrows}
\usepackage{cite}
\usepackage{subfig}
\usepackage{color}
\usepackage[font=footnotesize]{caption}
\usepackage{amssymb}
\usepackage{tabulary}
\usepackage{booktabs}
\usepackage{xcolor}
\usepackage{multicol}
\usepackage{bbm} 
\usepackage{dsfont}
\usepackage{macros}

\tikzstyle{int}=[draw, fill=white!20, minimum size=2em]
\tikzstyle{init} = [pin edge={to-,thin,black}]

\IEEEoverridecommandlockouts                          
\begin{document}
\title{Wireless Caching Helper System with Heterogeneous Traffic and Random Availability}
%\vspace{-5mm}
\author{
	\IEEEauthorblockN{Ioannis Avgouleas, Nikolaos Pappas, and Vangelis Angelakis}\\
%	\IEEEauthorblockA{Department of Science and Technology, Link{\"o}ping University, 60174 Norrk{\"o}ping, Sweden\\
%		Emails: \{ioannis.avgouleas, nikolaos.pappas, vangelis.angelakis\}@liu.se  }
}

\maketitle
\thispagestyle{plain}
\pagestyle{plain}

\IEEEpeerreviewmaketitle 

\begin{abstract}
Multimedia content streaming from Internet-based sources emerges as one of the most high demanded services by wireless users.
In order to alleviate excessive traffic due to multimedia content transmission, many architectures (e.g., small cells, femtocells, etc.) have been proposed to offload such traffic to the nearest (or strongest) access point also called ``helper". 
The deployment of more helpers is not necessarily beneficial due to their potential of increasing interference.
In this work, we evaluate a wireless system in which we distinguish between cachable and non-cachable traffic. 
More specifically, we consider a general system in which a wireless user with limited cache storage requests cachable content from a data center that can be directly accessed through a base station. 
The user can be assisted by a pair of wireless helpers that exchange non-cachable content as well. 
Packets arrive at the queue of the source helper in bursts.
Each helper has its own cache to assist the user's requests for cachable content. 
Files not available from the helpers are transmitted by the base station.
We analyze the system throughput and the delay experienced by the user and show how they are affected by the packet arrival rate at the source helper, the availability of caching helpers, the caches' parameters, and the user's request rate by means of numerical results.
\end{abstract}	
\section{Introduction}
Wireless video has been one of the main generators of wireless data traffic. 
It is expected to originate 75\% of the global mobile traffic by 2020 \cite{Cisco} and inevitably contribute to networks' congestion and delays.  
One of the most promising technologies to cope with such issues is caching popular files in helper nodes that constitute a wireless distributed caching network that assists base stations by handling requests for popular files \cite{Paschos_IEEE_Comm_Mag, FemToCaching}. 

Wireless caching helpers can store a number of popular files and  transmit them to the requesting users more efficiently, considering that helpers have been deployed in such a way that the wireless channel between helpers and users is better than the one between users and base stations.
%by exploiting the better characteristics of the wireless channel between helpers and users. %\cite{Pappas_Zheng_Access_2019}. 
Wireless networks with caching capabilities can significantly reduce cellular traffic and delay as well as simultaneously increase throughput \cite{Paschos_JSAC, Fundamental_Limits_of_Caching}.

In this paper, we study a wireless system in which traffic is distinguished in cachable and non-cachable.
A user with limited cache storage requests cachable content from a data center using a base station which has direct access to it through a backhaul link. Two wireless nodes within the proximity of the user exchange non-cachable content and have limited cache storage. 
Therefore, they can act as caching helpers for the user by serving its requests for cachable content.
Files not available at the helpers can be fetched by data center through the base station.
%We assume the content placement is given and hierarchical i.e., when the user requests for a file that is not stored in his most popular files, it first probes the closest caching helper which stores the next most popular files of the library. 
%If this probe fails, then the second caching helper is probed. If the latter misses the file, then it can be found in the data center. 
Additionally, the source helper is equipped with a queue whose role is to save the excessive traffic with the intention of transmitting it to the destination helper in a subsequent time slot. Packets arrive in bursts. 
Concerning caching, we assume the content placement is given and hierarchical.

\subsection{Related Work}

Various content placement strategies have been studied in scientific literature e.g., caching the most popular content everywhere \cite{Paschos_IEEE_Comm_Mag}, probabilistic caching \cite{Prob_Caching_Cache_Hit_vs_Througput_Optimal,Delay_Geo_Caching_2tier_HetNets}, cooperative caching \cite{Pappas_Zheng_Access_2019, Cooperation_Caching_for_HetNets_2017, Cooperative_Caching_Placement_in_Cached_D2D_Underlaid_Cellular_Nets, Chen_CooperativeCaching_TWC, FD_communications_2018}, or caching based on location e.g., geographical caching \cite{D2D_vs_Small_Cell_Caching}.

Additionally, several different performance metrics have been considered. 
In earlier studies of wireless caching, cache hit probability (or ratio) \cite{D2D_vs_Small_Cell_Caching}, and the density of successful receptions or cache-server requests \cite{D2D_vs_Small_Cell_Caching, Prob_Caching_Cache_Hit_vs_Througput_Optimal} have been commonly investigated as a means of evaluating the performance of wireless caching systems.
Furthermore, there are several studies regarding energy efficiency or consumption of the different caching schemes \cite{D2D_vs_Small_Cell_Caching, EE_Downlink_Caching_at_BS, EE_Wireless_Caching_D2D_Cooperative, Cached_D2D_Communications_Offloading_Gain_and_Energy_Cost} as well as taking into account the traffic load of the wireless links \cite{Optimal_Caching_Placement_for_D2D, Spatially_Caching_for_D2D}. 
Methods that reduce traffic load by  optimizing the offloading probability or gain can be found in \cite{EdgeCaching_D2D_offloading,Optimal_Caching_and_Scheduling_for_Cache_enabled_D2D_communications,CachingDiversity}.

More recently, a considerable amount of research works analyze wireless caching systems by considering throughput\cite{Pappas_Zheng_Access_2019, Delay_Geo_Caching_2tier_HetNets} and/or delay \cite{OnlineCaching}. Regarding the latter, the majority of research works cope with mitigating the backhaul or transmission delay under the assumption that traffic or requests are saturated. However, there are works that take into account stochastic arrivals of requests at different nodes \cite{Single_Bottleneck_Caching_Networks_Analysis, Adaptive_Video_Steaming_with_Multiple_Helpers}.

Caching has been applied to several  different network realizations e.g., FemtoCaching \cite{FemToCaching}
in which the so-called Femto Base Station (FBS) serve a group of dedicated users with random content requests while simultaneously the non-dedicated users might be served with delay due to cache misses or no FBS availability. The coded/uncoded cached contents are stored in multiple small cells, the so-called femtocells. Given the file requests distribution and the cache size of each femtocell, the content placement is studied such that the downloading time is minimized.

The advent of vehicular networks necessitates the use of caches in order to reduce the latency of content streaming and increase the offered Quality of Service (QoS) \cite{VehNets_OnDemandStreaming, VehNets_qlearning}. Supporting vehicle-to-everything connections urges the exploration of alternative data routing protocols in order to avoid incuring excessive end-to-end delay and backhaul resources allocation. 
On the contrary, moving computational and storage resources to the mobile edge computing seems encouraging \cite{MEC_survey, CachingScheme_MEC, CachingChallenges}. This can be done e.g., by employing a new paradigm known as local area data network \cite{VehNets_LowLatencyRadioAccess}, or other advances in radio access networks (RANs) for Internet of Things (IoT) \cite{Advances_Edge_IoT}.

Many contemporary works consider to jointly optimize the problems of content caching (or placement), computing, and allocating radio resources. They usually consider and solve separately these important issues by formulating the computation offloading or content caching as convex optimization problems with different metrics e.g., service latency, network capacity, backhaul rate etc. \cite{JointOptimization1,JointOptimization2}. Works that simultaneously address the aforementioned problems together and propose a joint optimization solution for the fog-enabled IoT or cloud RANs (C-RANs) can be found in \cite{Fog_JointCachingComputing} and \cite{CRAN_ContentPlacement}, respectively. 

For some applications e.g., broadcast or multicast applications, single transmissions from the base station to more than one user is useful. The authors in \cite{SmartGrid_caching} propose a content caching and distribution scheme for smart grid enabled heterogeneous networks, in which each popular file is stored in multiple service nodes with energy harvesting capabilities. The optimization of the total on-grid power consumption, the user association scheme, and the radio resource allocation improves the reliability and performance of the wireless access network. 
The evolution of 5G mobile networks is going to incorporate cloud computing technologies. The authors in \cite{CaaS} propose the concept of ''Caching-as-a-Service'' (CaaS) based on C-RANS as a means to cache anything, anytime, and anywhere in the cloud-based 5G mobile networks with the intention of satisfying user demands from any service location with high QoS. Furthermore, they discuss the technical details of virtualization, optimization, applications, and services of CaaS in 5G mobile networks.

A key distinction among research papers in wireless caching is the assumption regarding the availability of caching helpers. Many papers consider that caching helpers can serve users requests whenever the requested file is cached while others adopt the more realistic assumption that caching helpers might be unable to assist user requests when, for example, serve other users of interest \cite{FemToCaching, D2D_vs_Small_Cell_Caching}. 
%The bursty arrivals at caching helpers can affect its performance e.g., the response time for its dedicated and non-dedicated users and, hence, the throughput and delay of the wireless system.
To the best of our knowledge, the proposed wireless caching model has not been studied in literature. The work of \cite{Pappas_Zheng_Access_2019} analyzes a wireless network with one caching helper that does not consider hierarchical caching.

\subsection{Contribution}
%We analyze the network-wide throughput and delay based on the assumption that wireless nodes have random access to the channel. First, we characterize the network-wide throughput concerning the case in which the queue at the source helper is stable as well as unstable. Afterwards, we optimize the maximum weighted network throughput subject to source's queue being stable. Subsequently, we characterize the average delay seen by the user in case of a local miss and, finally, we provide numerical results.  

In this paper, we study a wireless system in which traffic is distinguished in cachable and non-cachable.
A user with limited cache storage requests cachable content from a data center connected to a base station through a backhaul link. Two wireless nodes within the user’s proximity exchange non-cachable files and have limited cache storage. 
Therefore, they can act as caching helpers for the user by serving its requests for cachable content.
Files not available at the helpers can be transmitted by the base station.
The source helper is equipped with an infinite queue whose role is to save packets intended for the destination helper for transmission in a subsequent time slot. Packets arrive in bursts.

We analyze the system throughput by assuming that the wireless helper nodes have random access to the channel and there is no coordination between them. 
First, we characterize the system throughput concerning the case in which the queue at the source helper is stable as well as unstable. 
Moreover, we formulate a mathematical optimization problem to optimize the probabilities by which the helpers assist the user to maximize the system throughput.
Subsequently, we characterize the average delay experienced by the user from the time of a local cache miss until it receives the requested file. 
Finally, we provide numerical results to show how the packet arrival rate at the source helper, the availability of caching helpers, random access to the channel, caching parameters, and the user's request rate affect the system throughput and delay.
 
\subsection{Organization of the paper}  
In Section \ref{Sec:SystemModel}, we present the system model comprising the network, the caching, the transmission, and the physical layer model. Section \ref{Sec:Throughput} provides the analytical derivation of throughput for the cases of stable and the unstable queue at the source helper. The average delay performance is given in Section \ref{Sec:Delay}. In Section \ref{Sec:Results}, we numerically evaluate our theoretical analysis of the previous sections. 
Finally, Section \ref{Sec:Conclusion} concludes our research work.
\section{System Model}\label{Sec:SystemModel}
\subsection{Network Model}

\begin{figure}[t!]	
	\centering
	\includegraphics[width=1\linewidth]{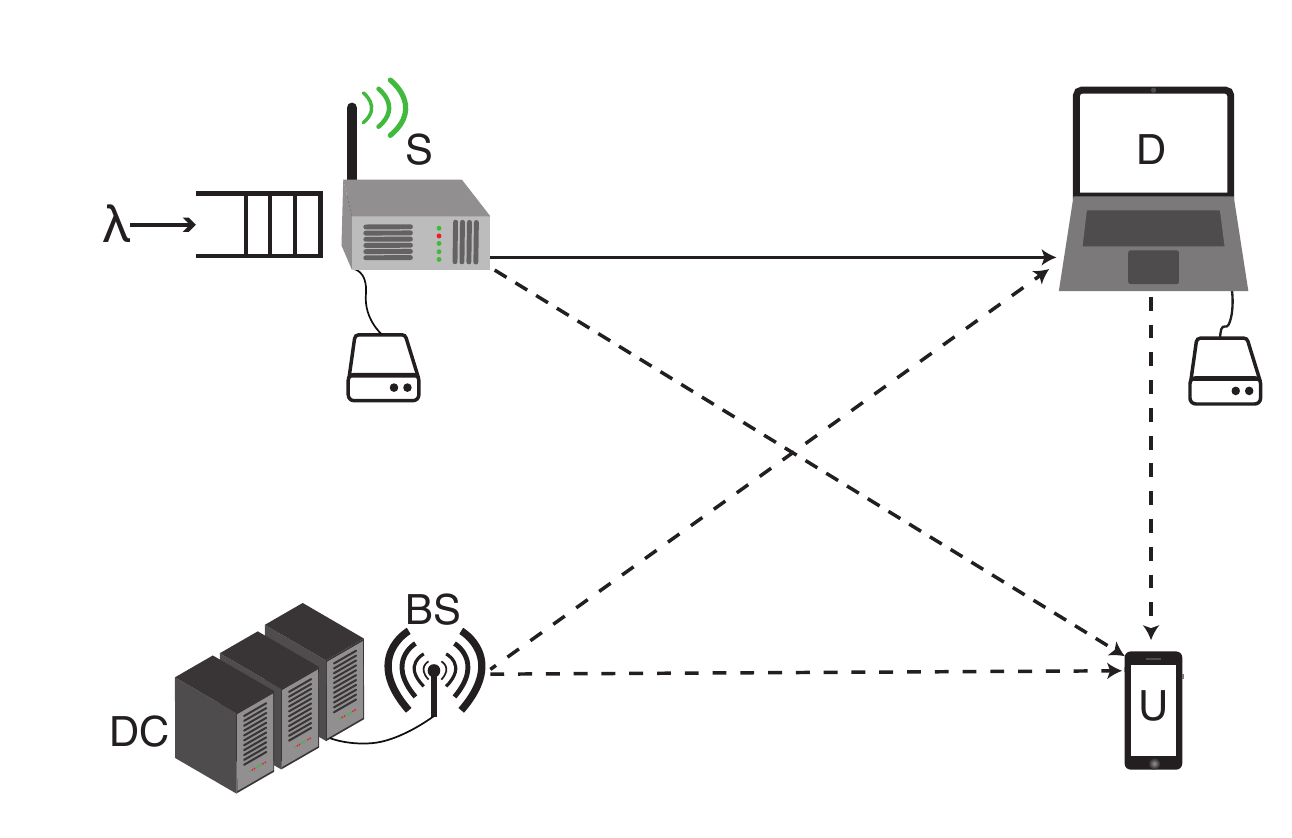}
%	\caption{The wireless network we analyze in our system model. Packets randomly arrive to the queue of node $S$  who serves its dedicated user $D$ with average arrival rate $\lambda$.	Both $S$ and $D$ have their own cache to serve user $U$'s need for multimedia content. 	User $U$ is also within proximity of the data center $DC$ through the base station $BS$. It is preferable to $U$ when it is served by $D$ or $S$, in case of $D$'s failure, since the $DC-U$ link can be problematic.}
%	\caption{The wireless network we analyze in our system model: user $U$ requests multimedia content from the data center $DC$ which can be directly accessed through the base station $BS$. The pair $S$-$D$ is a separate network within $U$'s proximity. Packets randomly arrive with average arrival rate $\lambda$ to the queue of node $S$ who serves its dedicated user $D$ with its intended traffic. Both $S$ and $D$ have their own caches to serve user $U$'s requests. It is preferable to $U$ when it is served by $D$ or $S$, in case of $D$'s failure, since the $DC-U$ link can be problematic.}
\caption{The wireless network we analyze in our system model: user $U$ requests cachable content from the data center $DC$ which can be directly accessed through the base station $BS$. The pair $S$-$D$ exchanges non-cachable content and falls within $U$'s proximity. Packets randomly arrive with average arrival rate $\lambda$ to the queue of node $S$ who serves node $D$ with non-cachable traffic. Both $S$ and $D$ have their own caches to serve user $U$'s requests for cachable content. 
It is preferable to $U$ when it is served by $D$ or $S$, in case of $D$'s cache miss or failure, since the link between $BS$ and $U$ can be problematic.}
\label{figure:SystemModel}		
\end{figure}

%\begin{itemize}
We consider a network system with four wireless nodes: a pair of caching helpers $S$ and $D$, a random user $U$ within the coverage of the helpers and a base station ($BS$) node connected to a datacenter ($DC$) through a backhaul link as depicted in Fig. \ref{figure:SystemModel}. We consider slotted time and that a packet transmission takes one time slot.

Helper $S$ is equipped with an infinite queue and the packet arrivals in its queue follow a Bernoulli process with average arrival rate $\lambda$. It transmits packets to the destination helper $D$.  
In each time slot, user $U$ requests for a file in its own cache. In case $U$'s cache miss, which happens with probability $\qu$, it requests the file from external resources i.e., the caching helpers or the data center (through $BS$). The data center stores the whole library and, hence, every file that $U$ may request.

Requesting a file directly from the $BS$ is not necessarily the best policy since there might be congestion at the link connecting $BS$ and $U$. Consequently, limited throughput or increased delay might be experienced using this link instead of requesting the file from one of the caching helpers. Moreover, the $BS$ is not always available to help $U$; this happens with probability $\alpha$ in each time slot. Therefore, it is preferable to $U$ when it is served by the caching helpers. 

\begin{figure*}	
	\centering
	\includegraphics[width=0.9\linewidth]{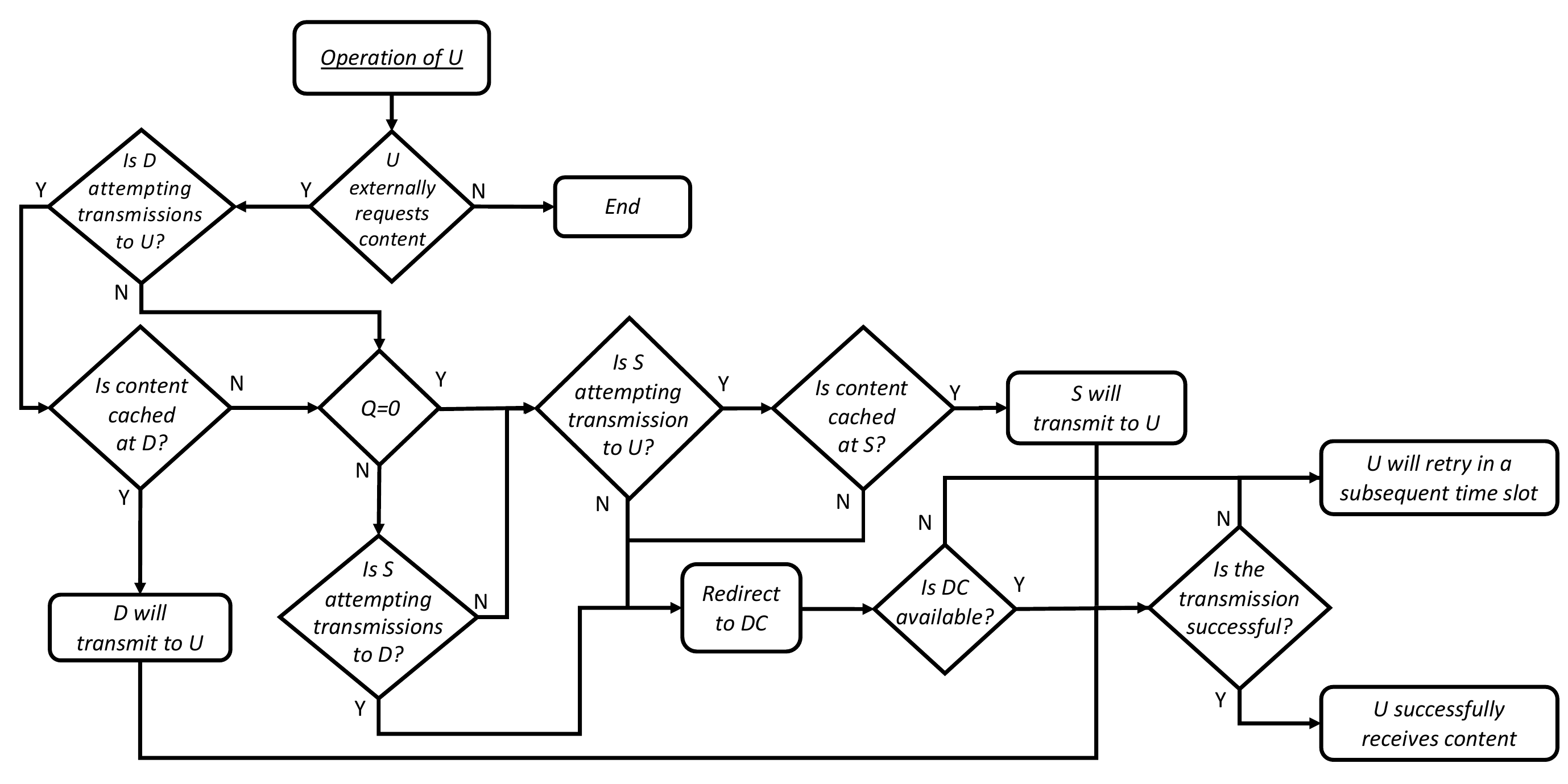}
	\caption{Operation of $U$ in the described protocol.}
	\label{Fig:operationU}		
\end{figure*}

\begin{figure*}	
	\centering
	\includegraphics[width=0.9\linewidth]{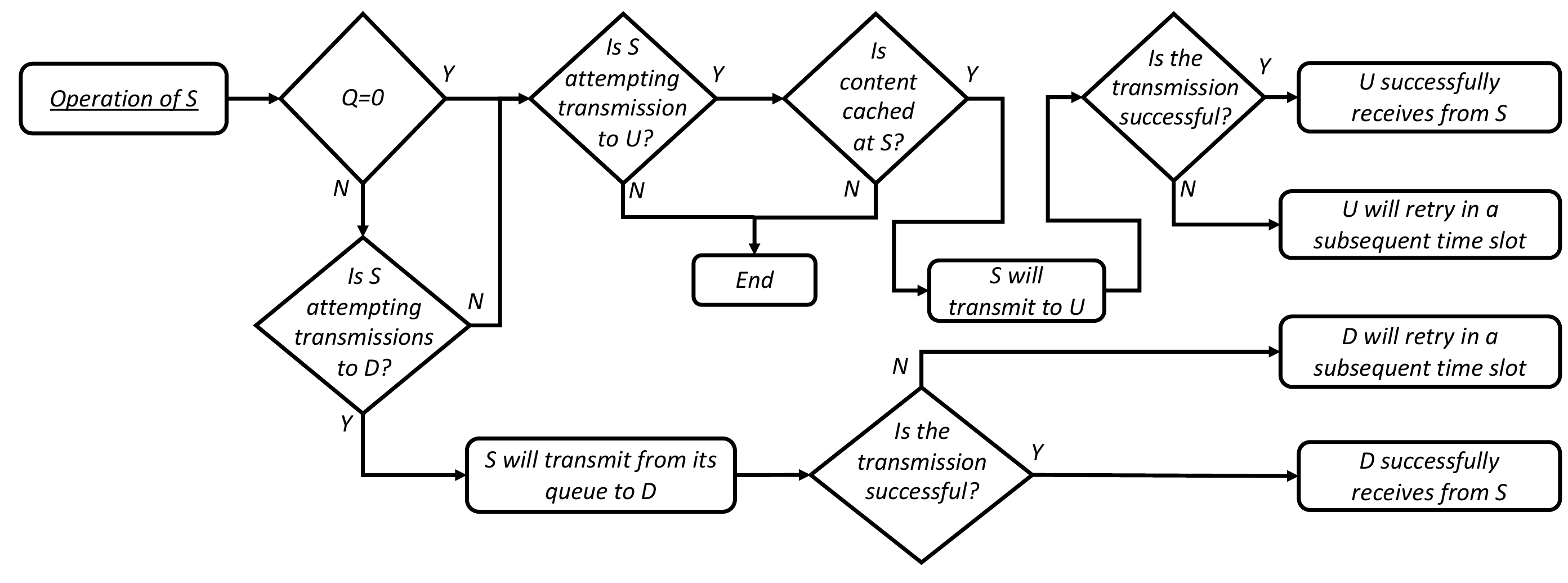}
	\caption{Operation of $S$ when user $U$ requests a file $f$ from external resources.}
	\label{Fig:operationS}		
\end{figure*}

\begin{figure*}	
	\centering
	\includegraphics[width=0.9\linewidth]{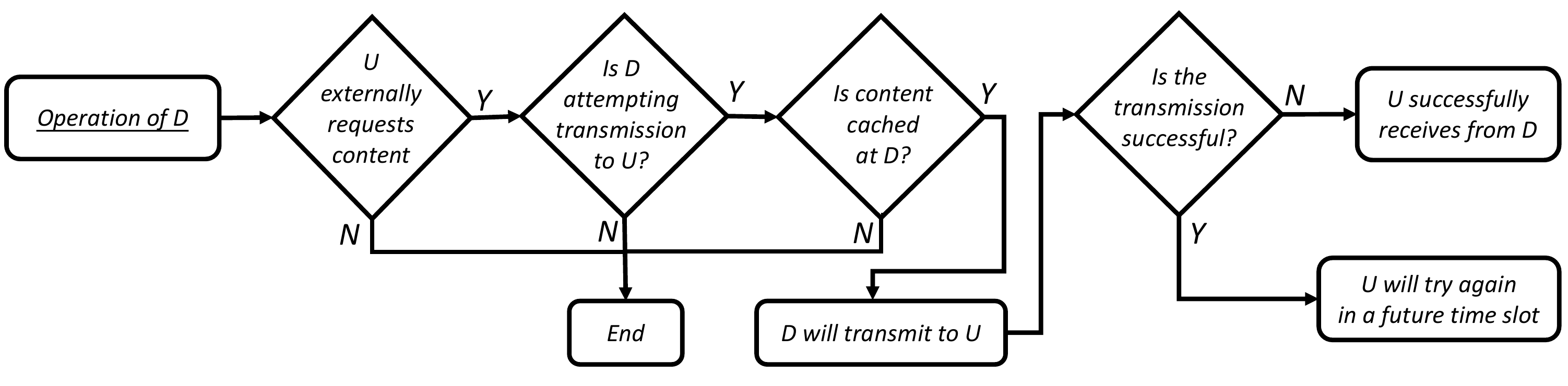}
	\caption{Operation of $D$ when user $U$ requests a file $f$ from external resources.}
	\label{Fig:operationD}		
\end{figure*}

The flowchart of user $U$'s operation with respect to its request content search is shown in Fig. \ref{Fig:operationU}. 
The operation of caching helpers $S$ and $D$, represented as flowcharts, are depicted in Figs. \ref{Fig:operationS} and \ref{Fig:operationD}, respectively.

\subsection{Cache Placement and Access}\label{Sec:CachePolicy}
We assume the content placement is given and hierarchical i.e., when the user node requests for a file that is not stored in its most popular files, it first probes the closest caching helper which stores the next most popular files. If this probe fails, then the second caching helper is probed for the requested file. If it also cache misses, then the file can be found in the data center. 
Additionally, the source helper is equipped with a queue whose role is to save the excessive non-cachable traffic with the intention of transmitting it to the destination helper in a subsequent time slot.

Furthermore, the user device $U$ and the caching helpers $D$ and $S$ have cache capacity to $M_U$, $M_D$, and $M_S$ files, respectively, and $M_U  \leq M_D \leq M_S$ holds.
We also consider the Collaborative Most Popular Content (CMPC) policy. According to CMPC, user $U$ stores the first $M_U$ most popular files in its own cache, helper $D$ stores the next most $M_D$ popular files, and $S$ stores the next most $M_S$ popular files.
Following CMPC requires exchange of information among devices e.g., the cache size of each device and the content placement in each device. We assume that this information exchange is negligible.

\subsection{Transmission Model}
In each time slot, $S$ will attempt transmissions of non-cachable content to $D$ with probability $\qs$ (if its' queue is not empty), and is available for $U$ with \pro $1-\qs$. 
We assume that the caching helpers assist $U$ only when specific conditions apply: $D$ will attempt transmission to $U$ with probability $\qd$, and $S$ will help $U$ only when it is not transmitting to $D$.
When the source caching helper $S$ is transmitting to helper $D$ and the user $U$ requests a file from external resources, then $U$ can be served by $D$ or by $DC$.
In that case, there are two parallel transmissions one from $S$ to $D$, and one from $D$ (or $DC$ respectively) to $U$.
If the caching helper $S$ is available for $U$, then there is no parallel transmissions since only one of $S$, $D$, or $DC$ can help $U$ at the same time slot.

Regarding $DC$, we model its availability with a \pro $\alpha$ to model the fact that it is not always available to serve $U$ due to serving other users or failure. 
If the $DC$ is always available to $U$, then $\alpha = 1$. On the other hand, if the $DC$ is not available for $U$, then $\alpha = 0$.
\begin{table}[t!] \caption{Notation table}
	\def\arraystretch{1.2}%  1 is the default
	\begin{center}
		\begin{tabular}{ | c  l |}
			\hline
			\textbf{Notation} & \textbf{Event} \\ \hline
			$\qu$   & $U$ requests file from external resources \\
			& ($U$ cache miss) \\ 
%			\hline
			$\phd$ & cache hit at $D$ (CMPC policy) \\ 
%			\hline
			$\phs$ & cache hit at $S$  (CMPC policy) \\ 
%			\hline
			$\qs$  & $S$ attempts to transmit to $D$ if $ Q \neq 0 $ \\
			& 	(or $S$ is available to serve $D$) \\ 
%			\hline
			$\qc$  & $S$ attempts to transmit  to $U$ \\
%			\hline
			$\qd$  & $D$ attempts to transmit  to $U$ \\ 
%			\hline
			$\A$ & the $DC$ is available to serve $U$ \\ 
%			\hline
			$\PSNR{i}{j} $ & success probability of link $i \rightarrow j$, \\
			& when $i$ is transmitting \\ 
%			\hline
			$\PSINR{i}{j}{k} $ & success probability of link $i \rightarrow j$,  \\
			& when $i$ and $k$ are transmitting \\ 
%			\hline
			$\prob( i \Rightarrow j) $& queued node $i$ attempts to transmit to $j$ \\ 
			\hline  
		\end{tabular}
	\end{center}
	\label{table:probabilities}
\end{table}
We summarize the aforementioned events and notation in Table \ref{table:probabilities}. 
Additionally, the operation of $U$, $S$, and $D$ as flowcharts can be found in Figs. \ref{Fig:operationU}-\ref{Fig:operationD}.

\subsection{Physical Layer Model}
We assume Rayleigh fading for the wireless channel and that a packet transmitted by node $i$ is successfully received by node $j$ if and only if $SINR(i,j) \geq \gamma_j$, where $\gamma_j$ is a threshold characteristic of node $j$. 
Therefore, the received power at node $j$ when $i$ transmits is $P_{rx}(i,j) = A(i,j)h(i,j)$, where $A(i,j)$ is exponentially distributed and the received power factor is:  
\[ h(i,j) = \frac{P_{tx}(i)}{r(i,j)^{p}},\]
%$h(i,j) =P_{tx}(i)r(i,j)^{-p}, $
where $P_{tx}(i)$ is the transmit power of node $i$, $r(i,j)$ is the distance in $m$ between node $i$ and node $j$, and $p \in [2,6]$ is the path-loss exponent.

By assuming perfect self-interference cancellation, the success probability of link $i \rightarrow j$, with $\mathcal{T}$ denoting the set of transmitting nodes, is given by \cite{Pappas_FD_TWC}:
%\[ \PSINR{i}{i}{\mathcal{T}} = exp\Big(  - \frac{\gamma_j n_j}{v(i,j)h(i,j)} \Big) (1 + \gamma_j(r(i,j)^\alpha g))^{-m} \times \prod_{k \in \mathcal{T}\setminus \{i,j\}  } \Big( 1 + \gamma_j \frac{v(k,j)h(k,j)}{v(i,j)h(i,j)}\Big)^{-1}, \]
\begin{align*}
\scriptsize
\PSINR{i}{j}{\mathcal{T}} &= exp\Big(-\frac{\gamma_j n_j}{v(i,j)h(i,j)}\Big)
\times \\
& \times  
\prod_{k \in \mathcal{T} \setminus \{ i,j \}} \Big( 1 + \frac{\gamma_{j} v(k,j)h(k,j)}{v(i,j)h(i,j)} \Big)^{-1},
\end{align*}

where  $v(i,j)$ is the parameter of the Rayleigh fading r.v., and $n_j$ is the noise power at receiver $j$.
%, and $m = \indicator(j \in \transmitters)$ with $\indicator(.)$ denoting the indicator function.
%For the sake of simplicity, we also use the notation $\PSNR{i}{j}$ to denote the success \pro of link $i \rightarrow j$ when node $i$ is the only transmitting node.
To simplify notation, we use $\PSNR{i}{j}$ to denote the success \pro of link $i \rightarrow j$ if node $i$ is the only transmitting node. 
\section{Throughput Analysis} \label{Sec:Throughput}
In this section, we analyze the throughput of the system depicted in Fig. \ref	{figure:SystemModel}. 
We are interested in the weighted sum of the throughput that helper $S$ provides along with the throughput realized by user $U$.
By denoting the former with $T_S$ and the latter with $T_U$, the weighted sum throughput is given by: 
\begin{equation}
wT_S + (1-w) T_U, \text{for } w \in [0,1].
\end{equation} 
The average service rate of caching helper $S$ is:
\begin{align} \nonumber
\mu ~=& ~q_s (1-\qu) \PSNR{S}{D} ~+~ q_s \qu \qd \phd \PSINR{S}{D}{D} ~+ \\ \nonumber
+ & ~q_s \qu (1 - \qd \phd)\alpha q_s \PSINR{S}{D}{DC} ~+ \\
+ & ~q_s \qu (1 - \qd \phd) (1-\alpha) \PSNR{S}{D}.
\end{align}   

As a corollary of the Loynes theorem \cite{loynes_1962}, we obtain that if the arrival and the service process of a queue are strictly jointly stationary and the queue's average arrival rate is less than the queue's average service rate, then the queue is stable. Thus, in our model, the queue at helper $S$ is stable if and only if $\lambda < \mu$.
Finite queueing delay is a ramification of a stable queue, and, hence, by adding the aforementioned constraint we can enforce finite queueing delay on our wireless system. 
Moreover, the stability at $S$ also implies that packets arriving at the queue will eventually be transmitted \cite{loynes_1962}. 

The throughput from $S$ to $D$, denoted by $T_S$, depends on the stability of the queue at $S$ and is $T_S = \lambda$ if the queue is stable or $T_S=\mu$ otherwise. Consequently:
\begin{equation} \label{eq:T_s}
	T_S = \mathds{1}(\lambda < \mu ) \lambda + \mathds{1}(\lambda \geq \mu) \mu, 
\end{equation}
with $\mathds{1}(.)$ denoting the indicator function.

The throughput realized by $U$, denoted by $T_U$, depends on whether the queue at $S$ is empty or not. The former happens with probability $\prob( Q=0) $ and the latter with probability $\prob( Q\neq0)$.
Therefore:
\begin{itemize}
	\item If the queue at $S$ is empty and $U$ requests a file from external resources.
	In this case, $U$ will be served: (i) by $D$ with probability $q_D$, or (ii) by $S$ with probability $q_C$ in case of $D$'s failure, or (iii) by the data center with probability $\alpha$ in case both helpers fail.
	\item If the queue at $S$ is non-empty and $U$ requests a file from external resources, then there are two cases: either (i)  helper $S$ attempts transmission to the destination helper $D$ (which happens with probability $q_S$) or (ii) helper $S$ is available to serve $U$. In the first case, $U$ will be served by $D$ with probability $q_D$ or by the data center in case $D$ fails to serve $U$. 
	In the second case, $U$ will be served by $D$ with probability $q_D$, or by $S$ with probability $q_C$ in case $D$ fails, or by the data center in case both helpers fail to serve $U$.
\end{itemize}
Considering all the details above, the throughput realized by user $U$ is:
\begin{align}\nonumber
	\scriptsize
	&T_U =\\\nonumber
	&= \pzero \qu \Big[ \qd \phd  \PSNR{D}{U} + (1 - \qd \phd)\qc \phs \PSNR{S}{U}\\\nonumber
	&+ (1 - \qd \phd)(1 - \qc \phs) \A \PSNR{DC}{U} \Big]+ \pnotZero \qu  \qs \times \\ \nonumber
	& \times  \Big[ \qd \phd \PSINR{D}{U}{S} + (1 - \qd \phd) \A \PSINR{DC}{U}{S} \Big] +\\ \nonumber
	& + \pnotZero \qu (1-\qs) \times \\ \nonumber
	& \times \Big[ \qd \phd \PSNR{D}{U} + (1 - \qd \phd)\qc \phs \PSNR{S}{U} + \\ 
	&+(1 - \qd \phd)(1-\qc \phs) \A \PSNR{DC}{U} \Big],
\end{align}
%\begin{small}
% \begin{equation*}
% T_u = \pzero \qu \Big[ \qd \phd  \PSNR{D}{U} + (1 - \qd \phd)\qc \phs \PSNR{S}{U}  +   
% \end{equation*}	
%\end{small} 
%\begin{equation*} 
%+ (1 - \qd \phd)(1 - \qc \phs) \A \PSNR{DC}{U} \Big] +
%\end{equation*}
%\begin{equation*} 
%+~ \pnotZero \qu  \qs \Big[ \qd \phd \PSINR{D}{U}{S}  +  (1 - \qd \phd) \A \PSINR{DC}{U}{S}   \Big] + 
%\end{equation*}
%\begin{equation*} 
%+~ \pnotZero \qu (1-\qs) \Big[ \qd \phd \PSNR{D}{U} + 
%\end{equation*}
%\begin{equation} \label{eq:T_u}
%+~ (1 - \qd \phd)\qc \phs \PSNR{S}{U} + (1 - \qd \phd)(1-\qc \phs) \A \PSNR{DC}{U} \Big],
%\end{equation}   
where we have to differentiate cases of stable/unstable queue due to different $\pzero$ and $\pnotZero$ for each case.
When the queue at $S$ is stable, the probability that $Q$ is not empty is given by:
$ \pnotZero = \lambda / \mu $.

In case the average arrival rate is greater than the average service rate i.e., $\lambda > \mu$, then the queue at $S$ is unstable and can be considered saturated. Consequently, we can apply a packet dropping policy to stabilize the system and the results for the stable queue can be still valid. 

If the queue at $S$ is unstable, the throughput realized by $U$ is: 
\begin{align} \nonumber
T'_U &= \qu  \qs \Big[\qd \phd \PSINR{D}{U}{S}  +  (1 - \qd \phd) \A \PSINR{DC}{U}{S}   \Big] + \\ \nonumber
+& \qu (1-\qs) \Big[ \qd \phd \PSNR{D}{U} + (1 - \qd \phd)\qc \phs \PSNR{S}{U} ~+ \\
+& (1 - \qd \phd)(1-\qc \phs) \A \PSNR{DC}{U} \Big] \label{eq:T'_u}
\end{align}    

We formulate the following mathematical optimization problem to optimize the probabilities $\qs,~\qc,$ and $\qd$ that maximize the weighed sum throughput when the queue at helper $S$ is stable:
\begin{subequations}
\begin{align}\label{problem:optimizationA}
\scriptsize
max.& ~w\lambda + (1-w)T_U\\
s.t.  & ~0 \leq \lambda < \mu\\ 
	    0 \leq &\qd, \qc, \qd \leq 1\label{problem:optimizationC}
\end{align}
\end{subequations}
The first constraint ensures the stability of the queue at helper $S$ and the second one defines the domain for the decision variables. To solve the aforementioned problem for the case in which the queue at $S$ is unstable, we have to drop the first constraint and replace the expressions for $\lambda$ and $T_U$ with the ones for $\mu$ and $T'_U$, respectively.
In Section \ref{Sec:Results}, we provide results for maximizing the weighted sum throughput for some practical scenarios.
\section{Delay Analysis}\label{Sec:Delay}
Delay experienced by users is another critical performance metric concerning wireless caching systems.
In this section, we study the delay that user $U$ experiences when requesting cachable content from external sources until that content is received.
Let $ P ( S \Rightarrow D ) = q_s P(Q \neq 0) $, $P ( S \not\Rightarrow D )  = 1 - P ( S \Rightarrow D )$  and $ \bar{q}_i = 1-q_i$. 

The average delay that user $U$ experiences to receive a file from external resources is:

\[  D_U = \phd \big \{  P ( S \Rightarrow D ) [ (1-\qd)D_{DC,1,D} + \]
\[ \qd \PSINR{D}{U}{S} + \qd \notPSINR{D}{U}{S} (1+D_D)] + \]
\[ +P ( S \not\Rightarrow D ) [ \qd  \PSNR{D}{U}+ \notqd  \phs D_{S_2} + \notqd \notphs D_{DC,0,D} ] \big \} +\]
\[  + \notphd \phs \times \]
\[ \times \big\{ P ( S \Rightarrow D ) [ \A \PSINR{DC}{U}{S} + (1-\A \PSINR{DC}{U}{S})(1+D_{S_1})] +   \]
\[  + P ( S \not\Rightarrow D ) [ \qc \PSNR{S}{U} + \qc \notPSNR{S}{U}  (1+D_{S_1}) + \notqc D_{DC,0,S} ]   \big\} + \]
\[  + \notphd \notphs \times \]
\[ \times \big \{ P ( S \Rightarrow D ) [ \A \PSINR{DC}{U}{S} + (1-\A \PSINR{DC}{U}{S})(1+D_{DC}) ] +   \]
\begin{equation}\label{eq:delay_U} + P ( S \not\Rightarrow D ) [ \A \PSNR{DC}{U} + (1-\A \PSNR{DC}{U}) (1+D_{DC}) ] \big \}, \end{equation}
where $D_{S_1}$ is the delay to receive the file from $S$ given $D$ misses it:
%\[ D_{S_1} = \big \{ \textit{"delay to receive $f$ from $S$} \]	
%\[ \textit{ given $f$ cache miss at $D$"}  \big\} = \]
\begin{equation} \label{eq:delay_DS1}
D_{S_1} = \qc \PSNR{S}{U} + \qc \notPSNR{S}{U} (1+D_{S_1}) +  \notqc D_{DC,0,S},
\end{equation}  
%, D_{S_2}, D_{DC},$ and $D_{DC, i, j}$ are given as follows:
and $D_{S_2}$ is the delay to receive the file from $S$ given $D$ caches it but does not attempt i.e., $\qd = 0$, transmissions to $U$:
%\[ D_{S_2} = \big \{\textit{"delay to receive $f$ from $S$ given $D$ caches $f$} \]	
%\[ \textit{but $D$ does not attempt $(\qd = 0)$ to transmit to $U$"} \big \} = \]
\begin{equation} \label{eq:delay_DS2}
D_{S_2} =  \qc \PSNR{S}{U} + ( 1 - \qc \PSNR{S}{U} ) (1 + D_D).
\end{equation}  
%\begin{equation}{\begin{multlined} D_{DC}  = P ( S \Rightarrow D )\big[ \A \PSINR{DC}{U}{S} + ( 1 - \A \PSINR{DC}{U}{S}) (1+D_{DC}) \big] +  \\
% + P ( S \not\Rightarrow D ) \big[ \A \PSNR{DC}{U} + ( 1 - \A \PSNR{DC}{U}) (1+D_{DC}) \big],\end{multlined}\end{equation} 
We also need to compute delay caused by the data center $DC$: 
\begin{align}  \label{eq:delay_DC}\nonumber
\scriptsize
	&D_{DC}  = \\\nonumber
	& P ( S \Rightarrow D )\big[ \A \PSINR{DC}{U}{S} + ( 1 - \A \PSINR{DC}{U}{S}) (1+D_{DC}) \big] + \\
	& + P ( S \not\Rightarrow D ) \big[ \A \PSNR{DC}{U} + ( 1 - \A \PSNR{DC}{U}) (1+D_{DC}) \big].
\end{align} 
Additionally, we need to calculate the following: 
\begin{align}
D_{DC, 0, S} =& \A \PSNR{DC}{U} + (1 - \A \PSNR{DC}{U}) (1 + D_{S_1}),  \label{eq:delay_DC0S} \\ 
D_{DC, 1, S} =&  \A \PSINR{DC}{U}{S} + (1 - \A \PSINR{DC}{U}{S})(1+D_{S_1}),\label{eq:delay_DC1S}\\ 
 D_{DC, 0, D} =& \A \PSNR{DC}{U} + (1 - \A \PSNR{DC}{U}) (1 + D_D), \label{eq:delay_DC0D} \\
D_{DC, 1, D} =& \A \PSINR{DC}{U}{S} + (1 - \A \PSINR{DC}{U}{S}) (1 + D_D),  \label{eq:delay_DC1D}  
\end{align}
and:
\begin{align}\nonumber
&D_D =  P ( S \Rightarrow D ) \times \\ \nonumber
& \times \{ \qd [\PSINR{D}{U}{S} + \notPSINR{D}{U}{S}(1+D_D)] + \notqd D_{DC,1,D} \} + \\ \nonumber
&+ P ( S \not\Rightarrow D ) \{ \qd [ \PSNR{D}{U} + \notPSNR{D}{U}(1+D_D)]  + \\ 
&+ \notqd (\phs D_{S_2}+\notphs D_{DC,0,D}) \}   \label{eq:delay_D} 
\end{align} 
As one can observe, (\ref{eq:delay_U}) - (\ref{eq:delay_D}) are recursively defined.
After some basic manipulations, (\ref{eq:delay_DC}) becomes:
\begin{equation}
D_{DC} = \big( \A \big[ P (S \Rightarrow D)(\PSINR{DC}{U}{S}-\PSNR{DC}{U}) + \PSNR{DC}{U} \big] \big)^{-1}.  
\end{equation}  

Assuming that $\qc \PSNR{S}{U} - \qc \neq 1$, (\ref{eq:delay_DS1}) becomes:
\begin{equation} \label{eq:delay_DS1_simplified}
D_{S_1} = \big( \qc\PSNR{S}{U} + \A (1-\qc)\PSNR{DC}{U} \big)^{-1} . 
\end{equation}

Assuming that $\qc \PSNR{S}{U} + (1-\qc)\A \PSNR{DC}{U}  \neq 0$ and using (\ref{eq:delay_DS1_simplified}), (\ref{eq:delay_DC0S}) and (\ref{eq:delay_DC1S}) become:
\begin{equation} 
D_{DC, 0, S} = 1 + \frac{1- \A \PSNR{DC}{U}}{ \qc \PSNR{S}{U} + (1-\qc)\A \PSNR{DC}{U}  },
\end{equation}

\begin{equation} 
D_{DC, 1, S} = 1 + \frac{1- \A \PSINR{DC}{U}{S}}{ \qc \PSNR{S}{U} + (1-\qc)\A \PSNR{DC}{U}  }.
\end{equation}

Using (\ref{eq:delay_DS2}), (\ref{eq:delay_DC0D}), (\ref{eq:delay_DC1D}) and applying the regenerative method \cite{Walrand}, we get:
\begin{align} \nonumber
 D_D &= \bar{q}_D P (S \not\Rightarrow D)\qc \phs \PSNR{S}{U} ~+ \\ \nonumber
 +& \bar{q}_D \alpha [ P (S \Rightarrow D)\PSINR{DC}{U}{S} + P (S \not\Rightarrow D)\bar{p}_{hS}\PSNR{DC}{U}  ] + \\ 
+& \qd [ \PSNR{D}{U} +  P (S \Rightarrow D) (\PSINR{D}{U}{S}-\PSNR{D}{U})].\label{eq:delay_D_D}
\end{align}

Substituting (\ref{eq:delay_D_D}) to (\ref{eq:delay_DS2}), (\ref{eq:delay_DC0D}), and (\ref{eq:delay_DC1D}) yields expressions for $D_{S_2}$,  $D_{DC, 0, D}$, and $ D_{DC, 1, D}$, respectively, that are functions of link success probabilities (see Table \ref{table:probabilities} and \ref{table:links_params}) and cache parameters (see Table \ref{table:caches_params}) only.

\section{Numerical Results}\label{Sec:Results}
In this section, we present numerical evaluations of the analysis in the previous sections.
The parameters we used for the wireless links between wireless nodes can be found in Table \ref{table:links_params}. 
The helpers apply the CMPC policy as described in Section \ref{Sec:CachePolicy}. 
We consider a finite content library of files, $\mathcal{F} = \{ f_1, ..., f_N \}$, to serve users requests.
For the sake of simplicity, we assume that all files have equal size and that access to cached files happens instantaneously.
The $i$-th most popular file is denoted as $f_i$, and the request probability of the $i$-th most popular file is given by: $	p_i =  \Omega / i^{\delta},$ where $\Omega = \big( \sum_{j=1}^N j^{-\delta} \big)^{-1}$ is the normalization factor and $\delta$ is the shape parameter of the Zipf law which determines the correlation of user requests.

Therefore, the \pro that user $U$ requests a file that is not located in its cache is:
\begin{equation} \label{eq:qu}  \qu = 1 - \sum\limits_{i=1}^{M_U} p_i,  \end{equation} 

the cache hit probabilities at the caching helper $D$ and $S$ are respectively given by:
\begin{equation}\label{eq:phd}
\phd = \sum_{i=M_U+1}^{M_U+M_D} p_i, 
\end{equation}
\begin{equation}\label{eq:phs}
\phs = \sum_{i=M_U+M_D+1}^{M_U+M_D+M_S} p_i.  \end{equation} 

In the following results, we study the \maxThr~which is defined as $T_w = wT_S + (1-w)T_U$ or $T'_w = wT_S + (1-w)T'_U$ when the queue at $S$ is stable  or unstable, respectively. The expressions for $T_S, T_U,$ and $T'_U$ are given by (\ref{eq:T_s})-(\ref{eq:T'_u}) in Section \ref{Sec:Throughput}. 
In order to maximize the weighted sum throughput, we solved the optimization problem (\ref{problem:optimizationA})-(\ref{problem:optimizationC}) using the Gurobi optimization solver and report the results.

\subsection{Maximum Weighted Sum Throughput vs. Average Arrival Rate $\lambda$}
\label{sec:Throughput_vs_lambda}
We consider a scenario where the wireless links parameters follow the values in Table \ref{table:links_params}.
The cache sizes and cache hit probabilities are set as per Table \ref{table:caches_params} for two different values for the variable $\delta$ of the standard Zipf law for the popularity distribution that the cached files follow.
\begin{table}[h] \caption{Wireless links parameters.} 
	\def\arraystretch{1.1}%  1 is the default
	\begin{center}
		\begin{tabular}{ | c  c |  } 
			\hline \textbf{Parameter} & \textbf{Value} \\ \hline
			\hline
			$P_{tx}(S)$ & $ 1 ~mW$ \\ 	
			%			\hline 
			$P_{tx}(D)$ & $0.5 ~mW$ \\ 
			%			\hline
			$P_{tx}(DC)$ & $ 10 ~mW$ \\ 
			%			\hline  
			$ n $ & $ 10^{-11} ~W $ \\ 	
			%			\hline
			$ r_{S \rightarrow D}$ & $ 50 ~m$ \\ 
			%			\hline
			$ r_{D  \rightarrow U}$ & $ 50 ~m $ \\ 
			%			\hline
			$ r_{S \rightarrow U}$ & $ 40 ~m $ \\ 
			%			\hline
			$ r_{DC \rightarrow U}$ & $ 80 ~m $ \\ 
			%			\hline
			$ r_{DC \rightarrow D}$ & $ 100 ~m $ \\ 
			%			\hline
			$ p $ & $ 4 $ \\ 
			\hline
		\end{tabular}  
		\begin{tabular}{ | c  c |  } 
			\hline \textbf{Parameter} & \textbf{Value} \\ \hline
			\hline
			$ P_{ S \rightarrow U} $	   & $ 0.903 $ \\ 
			%			\hline
			$ P_{ D \rightarrow U }$   	   & $ 0.607 $ \\ 
			%			\hline
			$ P_{ DC \rightarrow U }$     & $ 0.849 $ \\ 
			%			\hline
			$ P_{ DC \rightarrow U/S }$  & $ 0.115 $ \\ 
			%			\hline
			$ P_{ S \rightarrow D }$ 		& $ 0.779 $ \\ 
			%			\hline
			$ P_{ S \rightarrow D/D }$ 	   & $ 0.779 $ \\ 
			%			\hline
			$ P_{ S \rightarrow D/DC }$   & $ 0.223 $ \\ 
			%			\hline
			$ P_{ D \rightarrow U/S} $     & $ 0.029 $ \\ 
			%			\hline
			$ \gamma_1 $   & $ 0 ~dB $  \\ 
			%			\hline
			$ \gamma_2  $  & $ 0 ~dB $  \\ \hline
		\end{tabular} 
	\end{center}
	\label{table:links_params}
\end{table} 
\begin{table}[ht!] \caption{Caches parameters and hit probabilities for different values of $\delta$.}
	\def\arraystretch{1.2}%  1 is the default
	\begin{center}
		\begin{tabular}{ | c c |} 
			\hline \textbf{Parameter} & \textbf{Value} \\ \hline
			\hline
			$ M_U $   			  &   $200$   \\ 
			$ M_D $   			  &  $1000$  \\ 
			$ M_S $   			  &  $2000$  \\ 
			$ F     $   			 & $10000$ \\ 	\hline
		\end{tabular}
		\begin{tabular}{ | c  c  c |}
		\hline \textbf{Parameter} & \multicolumn{2}{c|}{\textbf{Value}} \\ \hline
		\hline 	\textbf{$\delta$}   & \multicolumn{1}{|c}{\textbf{$ 0.5$}} &  \multicolumn{1}{|c|}{ \textbf{$ 1.2$}} \\ \hline
		$ \qu $    & \multicolumn{1}{|c}{$ 0.865 $} & \multicolumn{1}{|c|}{$0.196$} \\ 
		$ \phd $ & \multicolumn{1}{|c}{$ 0.206 $}  & \multicolumn{1}{|c|}{$0.109$} \\ 
		$ \phs $  &  \multicolumn{1}{|c}{$ 0.221 $}  &  \multicolumn{1}{|c|}{$0.045$} \\ 
		\hline
		\end{tabular}
	\end{center} 
	\label{table:caches_params}
\end{table}

\begin{figure}[ht] 
%	\begin{center}
		\centering
		\subfloat[$ \delta = 0.5$]{
		\includegraphics[width=1.05\linewidth]{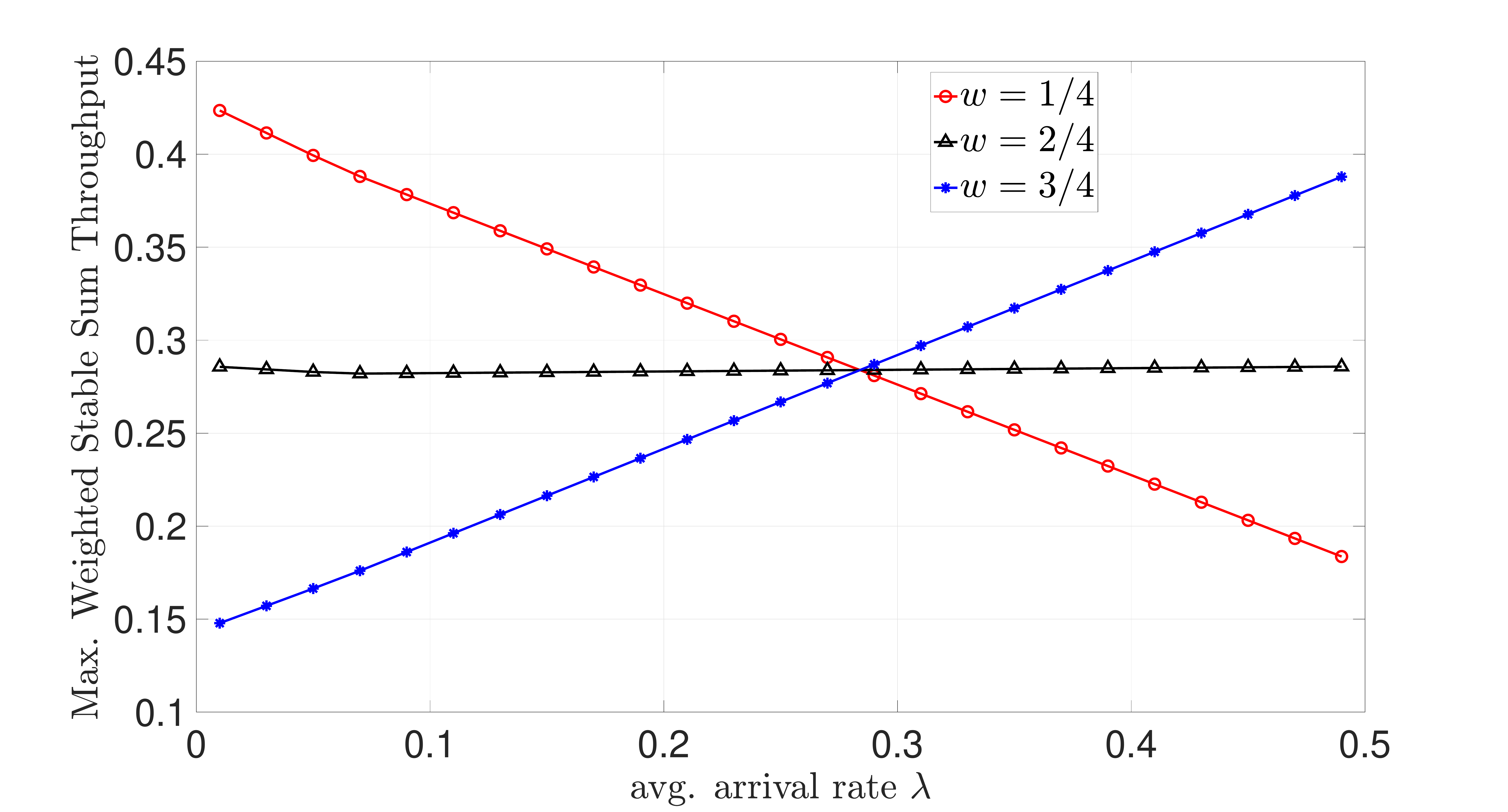}} 
		\hfill
		\subfloat[$ \delta = 1.2$]{
		\includegraphics[width=1.05\linewidth]{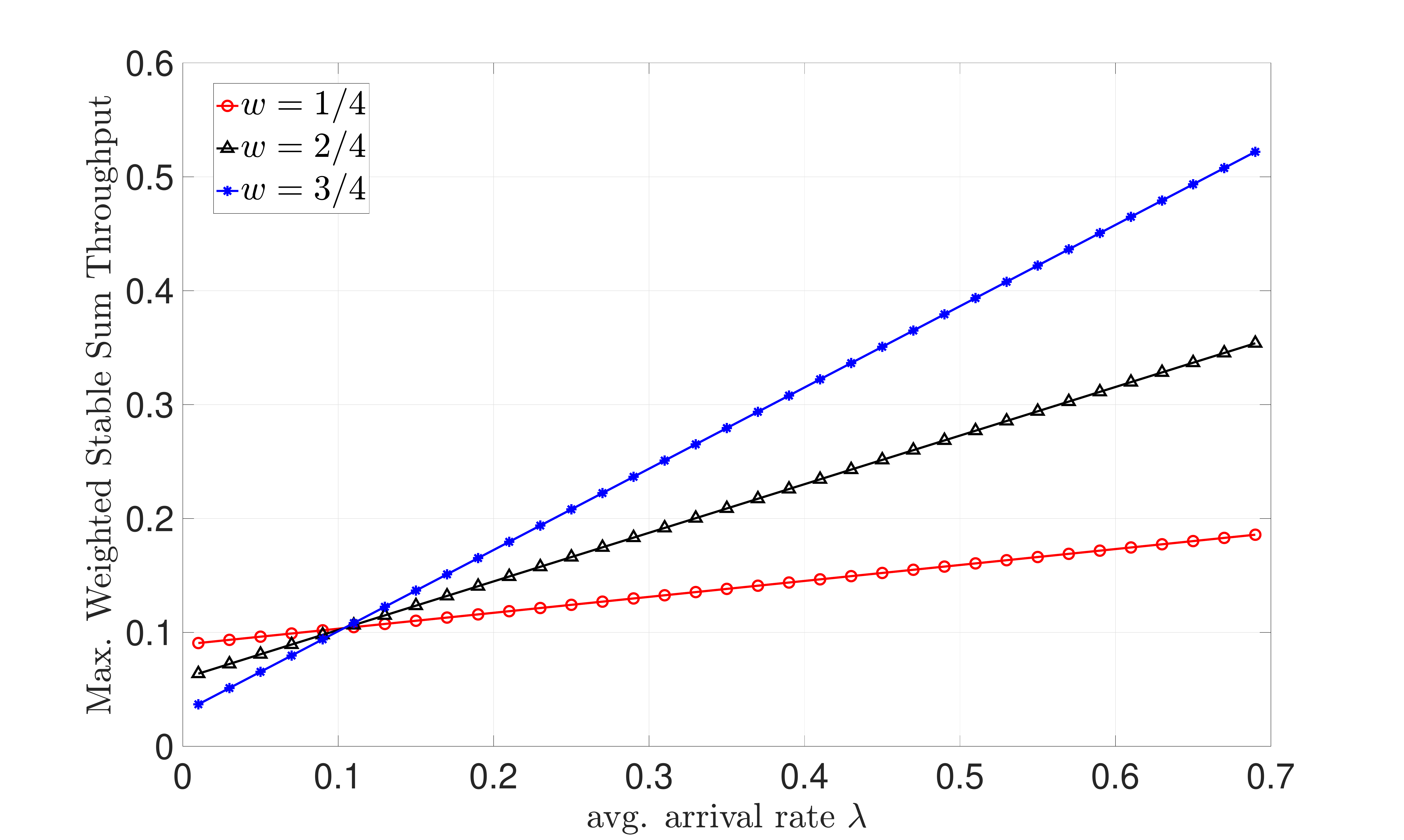}}
%	\end{center}
	\caption{The \maxThr ~vs. $\lambda$ for $\alpha=0.7$ and different values of $w$ when the queue at $S$ is stable for: (a) $ \delta = 0.5$ and (b) $  \delta = 1.2 $.}
	\label{Fig:stableThroughput_vs_lambda}
\end{figure}

In \figurename~\ref{Fig:stableThroughput_vs_lambda}, the \maxThr~versus the average arrival rate $\lambda$ at helper $S$ is presented for three different values of $w$ when the queue at $S$ is stable.
We chose: (i) $w=1/4$ as a representative case in which $T_U$ is more important than $T_S$, (ii) $w=2/4$ to equalize the importance of  $T_U$ and $T_S$, and (iii) $w=3/4$ to put more emphasis on the importance of $T_S$ versus $T_U$.

In case $w=1/4$, the \maxThr~is a decreasing function of $\lambda$ when $\delta=0.5$ (see Fig. \ref{Fig:stableThroughput_vs_lambda}(a)), but increasing when $\delta=1.2$  (see Fig. \ref{Fig:stableThroughput_vs_lambda}(b)). When $w=2/4$, the \maxThr~is almost constant for any value of $\lambda$ when $\delta=0.5$ and increases with $\lambda$  for $\delta=1.2$.
Regarding $w=3/4$, the \maxThr~is an increasing function of $\lambda$ for any $\delta$ value since $T_S$ clearly dominates $T_U$ in this case.

Furthermore, it is observed that the \maxThr~is achieved when $q^{*}_c =1$ for any value of $w$ and $\lambda$ when the queue at $S$ is stable (see Table \ref{table:stable_throughput}), but this is not the case when the queue is unstable i.e., the average arrival rate  $\lambda$ is greater than the average service rate $\mu$ (see Table \ref{table:unstable_throughput} for different values of $\delta$). 
In case the queue at $S$ is unstable, it is optimal for helper $S$ to avoid transmissions i.e., $q^{*}_c =0$, to $U$ when $\delta = 1.2$ for any values of $w$ and $\lambda$. However, this is not the case when $\delta=0.5$ since helper $S$ must always attempt transmissions to $U$ i.e., $q^{*}_c =1$, when $w \in \{ 1/4, 2/4 \}$ to achieve the \maxThr.

\begin{table}[h!] \caption{The values of $q^*_S, q^*_C, q^*_D$ for which the weighted sum throughput is maximized and the queue at $S$ is stable for $\alpha =0.7, M_U=200, M_D=1000,$ and $M_S =2000$.}
	\def\arraystretch{1.1}%  1 is the default
	\begin{center}
		\begin{tabular}{ | c || c | c | c | c || c | c | c | c | } \hline
			\multicolumn{1}{|c||}{} & \multicolumn{4}{c||}{$ \delta = 0.5$} & \multicolumn{4}{c|}{$\delta=1.2$} \\	
			\hline 
			\hline w & $T_w$  &  $q^*_S$ & $q^*_C$ &  $q^*_D$ & $T_w$  &  $q^*_S$ & $q^*_C$ &  $q^*_D$\\ \hline
			$ 1/4 $  & $ 0.423 $ & $0.029$ & $1$    &  $0$        & $ 0.187 $ & $ 0.988 $ & $1$& $1$ \\ \hline
			$ 2/4 $ & $ 0.286 $  & $0.978$ & $1$    &  $1$        & $ 0.358 $ & $ 0.985 $ & $1$& $1$\\ \hline
			$ 3/4 $ & $ 0.392 $  & $0.996$ & $1$   & $1$        & $0.536 $ & $ 0.999 $& $1$& $1$ \\ \hline
		\end{tabular}
		\\
%		\begin{tabular}{ | c || c | c | c | c | }
%			\hline  \multicolumn{5}{|c|}{$\delta = 1.2$} \\	\hline			
%%			\hline  & max weighted sum  & & & \\ 
%%			w & \textbf{stable throughput}  &  $q^*_S$ & $q^*_C$ &  $q^*_D$ \\ \hline
%			\hline w & $T_w$  &  $q^*_S$ & $q^*_C$ &  $q^*_D$ \\ \hline
%			$ 1/4 $  & $ 0.187 $ & $ 0.988 $ & $1$& $1$ \\ \hline
%			$ 2/4 $ & $ 0.358 $ & $ 0.985 $ & $1$& $1$ \\ \hline
%			$ 3/4 $ & $ 0.536 $ & $ 0.999 $& $1$& $1$ \\ \hline
%		\end{tabular}

		\quad
	\label{table:stable_throughput}
	\end{center}
\end{table}
\begin{table}[ht] \caption{The values of $q^*_S, q^*_C, q^*_D$ for which the weighted sum throughput is maximized and the queue at $S$ is unstable for $\alpha =0.7, M_U=200, M_D=1000,$ and $M_S =2000$.}
	\label{table:unstable_throughput}	
	\def\arraystretch{1.1} %  1 is the default
	\begin{center}
		\begin{tabular}{ | c || c | c | c | c || c | c | c | c | } \hline
			\multicolumn{1}{|c||}{} & \multicolumn{4}{c||}{$ \delta = 0.5$} & \multicolumn{4}{c|}{$\delta=1.2$} \\	
			\hline
			\hline w & $T'_w$  &  $q^*_S$ & $q^*_C$ &  $q^*_D$ & $T'_w$  &  $q^*_S$ & $q^*_C$ &  $q^*_D$ \\ \hline
			$ 1/4 $ & $ 0.430 $ & $0$ & $1$ &  $0$ & $0.189$  & $1$ & $0$ & $1$ \\ \hline
			$ 2/4 $ & $ 0.286 $ & $0$ & $1$&  $0$ & $0.363$  & $1$ & $0$ & $1$  \\ \hline
			$ 3/4 $ & $ 0.399 $ & $1$ & $0.024$ &  $1$ & $0.537$  & $1$ & $0$ & $1$ \\ \hline
		\end{tabular}
%		\begin{tabular}{ | c | c || c | c | c | } \hline 
%			\multicolumn{5}{|c|}{$\delta = 1.2$} \\ \hline 
%%			\hline  & max weighted sum  & & & \\
%%			w & \textbf{unstable throughput}  &  $q^*_S$ & $q^*_C$ &  $q^*_D$ \\ \hline
%%			\hline
%			\hline w & $T'_w$  &  $q^*_S$ & $q^*_C$ &  $q^*_D$ \\ \hline
%			$ 1/4 $   & $0.189$  & $1$ & $0$ & $1$  \\ \hline
%			$ 2/4 $  & $0.363$  & $1$ & $0$ & $1$ \\ \hline
%			$ 3/4 $  & $0.537$  & $1$ & $0$ & $1$ \\ \hline 
%		\end{tabular}
	\end{center}
\end{table}	
 
\subsection{Maximum Weighted Sum Throughput vs. Cache Size $M_U$}
In this section, we study how the cache size $M_U$ affects the \maxThr. 
Recall that $\qu$ decreases as $M_U$ increases. 
We consider two different values for $\delta$, same as previously, to examine how $\delta$ affects \maxThr~given different values for $M_U$. 

\begin{figure}[ht] 
	\centering
	\subfloat[$\delta = 0.5$]{
		\includegraphics[width=1.1\linewidth]{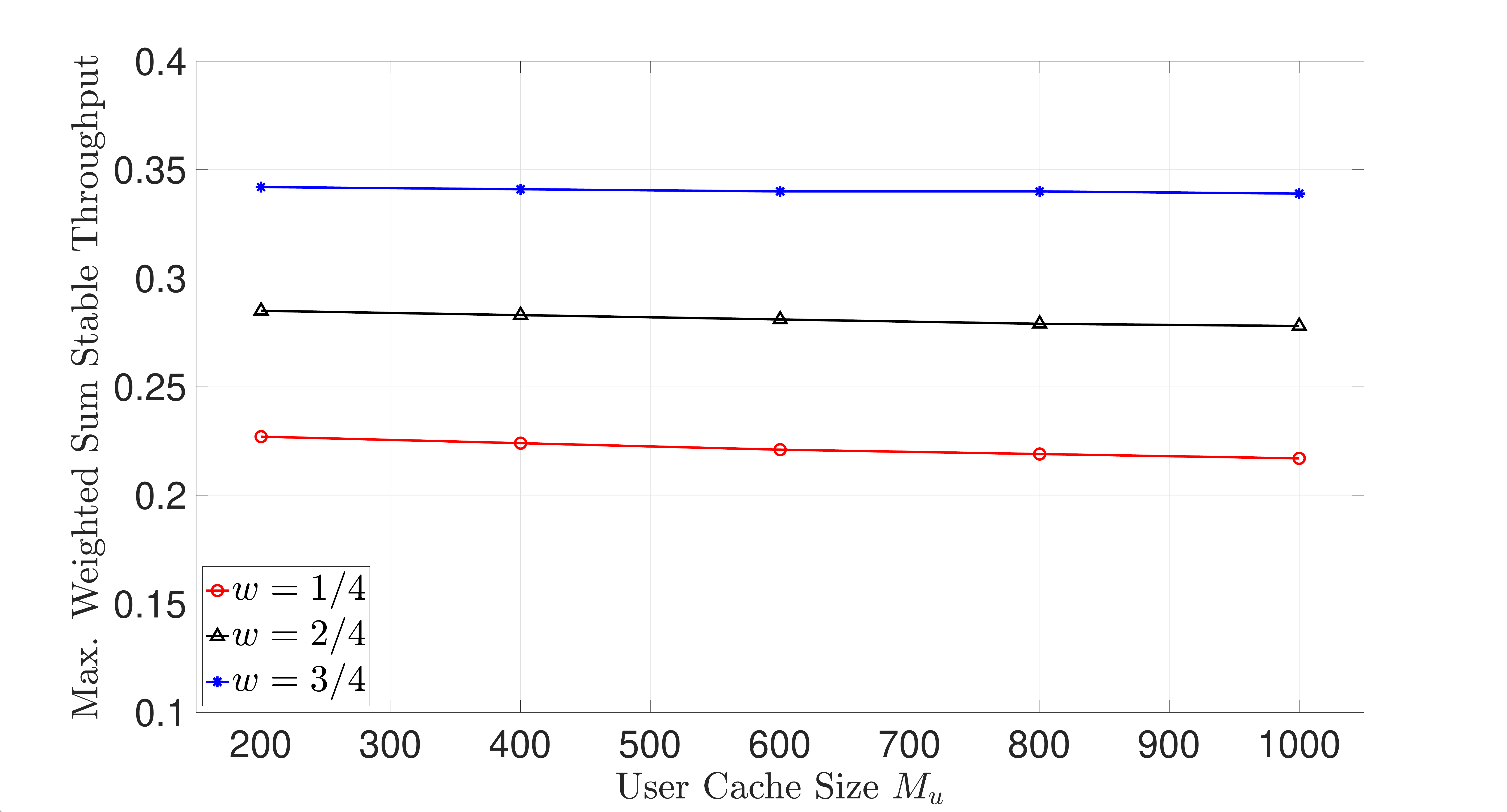}	}
	\hfill
	\subfloat[$\delta = 1.2$]{
			\includegraphics[width=1.1\linewidth]{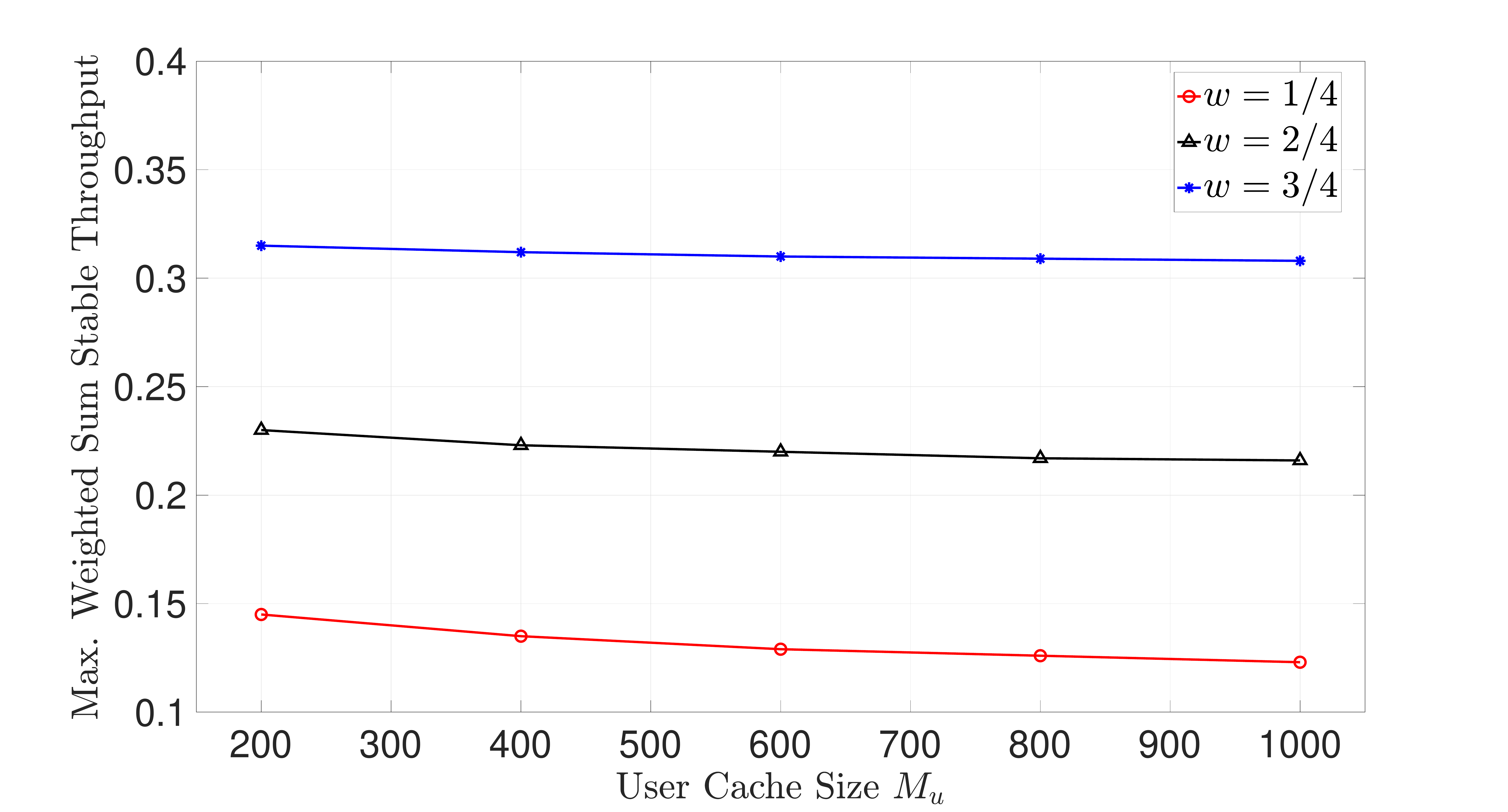}	}	
	\caption{The \maxThr ~vs. $M_U$ when the queue at $S$ is stable ($\lambda=0.4$) and $\alpha=0.7$ using different values of $w$ for: (a) $ \delta = 0.5$ and (b) $  \delta = 1.2 $.}
	\label{Fig:StableThroughput_vs_Mu}
\end{figure} 
\begin{figure}[ht]
	\centering
	\subfloat[$\delta = 0.5$]{
		\includegraphics[width=1.1\linewidth]{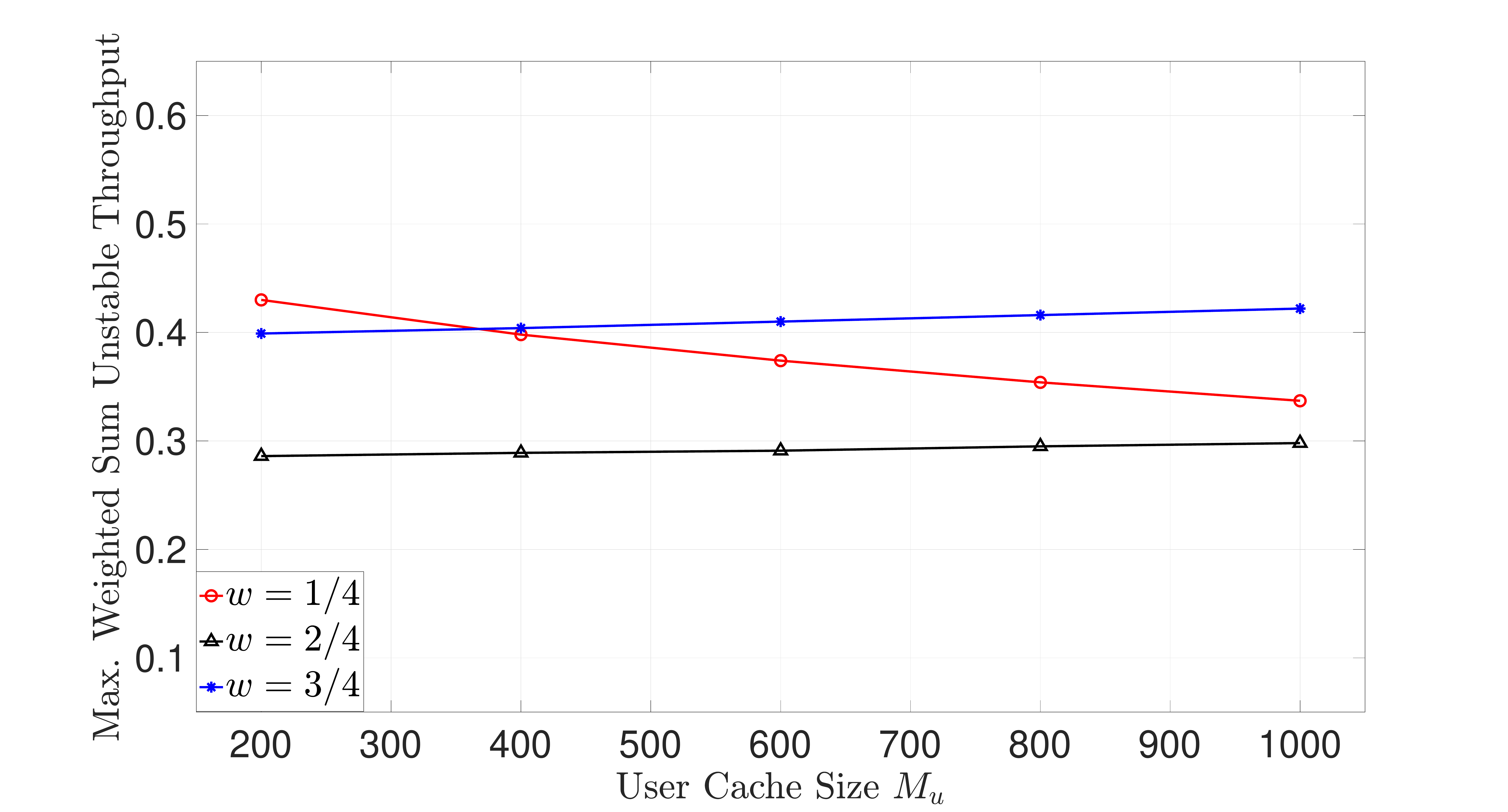}}
	\hfill
		\subfloat[$\delta = 1.2$]{
		\includegraphics[width=1.11\linewidth]{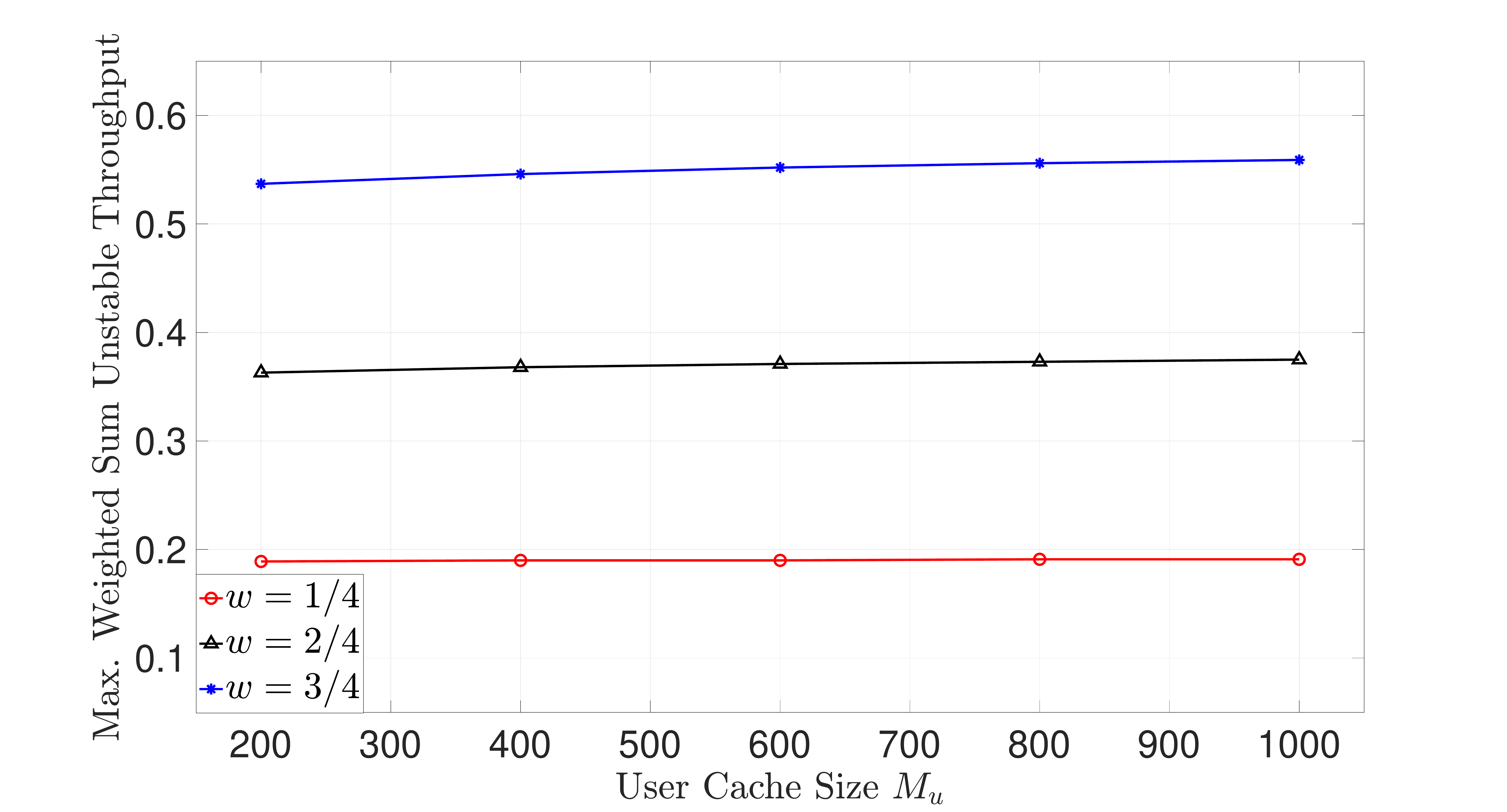}}		
	\caption{The \maxThr ~vs. $M_U$ for $\alpha=0.7$ using different values of $w$ when the queue at $S$ is unstable for: (a) $ \delta = 0.5$ and (b) $  \delta = 1.2 $.}
	\label{Fig:UnstableThroughput_vs_Mu}
\end{figure} 
In \figurename~\ref{Fig:StableThroughput_vs_Mu}, the \maxThr~versus $M_U$ is presented for $\alpha = 0.7, M_D = 1000, M_S = 2000$ and $\lambda = 0.4$ for which the queue at $S$ is stable. 
We observe that as the cache size at $U$ increases, the \maxThr~remains almost constant when $\delta = 0.5$ and slightly decreases when $\delta = 1.2$. This is expected since increasing cache size at $U$ results in fewer requests for files from external results. 
Moreover, the \maxThr~is higher when the value of $\delta$ is lower since, for a given cache size e.g. $M_U=200$, the probability of requesting content from external resources decreases as $\delta$ is increased.

In \figurename \ref{Fig:UnstableThroughput_vs_Mu}, the \maxThr~versus $M_U$ is presented for the same parameters as in the \figurename \ref{Fig:StableThroughput_vs_Mu} but unstable queue at $S$. 
The \maxThr~is an increasing function of $M_U$ for every value of $\delta$ when $w \in \{2/4, 3/4\}$. This is expected since, for these values of $w$, the throughput achieved by $D$ i.e., $T_S$, dominates $T_U$, and $T_S$ is increasing due to the decrease of requests to external content by $U$ (recall that as $M_U$ increases, $q_U$ decreases).
When $w=1/4$, the \maxThr~is almost constant ($\delta=1.2$) or decreases ($\delta=0.5$) as $M_U$ increases. 
The latter decrease can be attributed to the fact that $T_U$ i.e., the dominant term in the \maxThr, decreases as $\qu$ decreases (and $M_U$ increases). 

%In Fig. the \maxThr vs. $M_U$ is presented for $\lambda = 0.4$ and DC availability $\alpha=0.7$. The queue at $S$ is stable in this case. We used the values for $M_U$ described in Table [].
%As $M_U$ increases, the decrease in \maxThr is more intense for $\delta \in \{ 0.5, 0.75 \}$ . 
%In case $\delta =0.25$, there is a slight improvement of \maxThr as $M_U$ increases when $w = 1/4$. 
%This can be explained since increased $M_U$ yields decreased $\qu$ and, hence, the traffic inside the network gets decreased. 
%Less network traffic results in less interference and, thus, potentially increased $T_U$ i.e., throughput seen by user $U$.

The values of $q^*_S, q^*_C,$ and $q^*_D$ that achieve the \maxThr ~are given in Table \ref{table:StableThr_vs Mu} and \ref{table:UnstableThr_Mu} when the queue at $S$ is stable and unstable respectively. 

In case $\delta=0.5$ and the queue at $S$ is stable, the \maxThr~$T_w$ is achieved for $q^*_C =1$ and $q^*_D=1$  for every value of $M_U$ and $w$. This means that, for the aforementioned parameters, user $U$ should always be assisted by both $S$ and $D$ to achieve \maxThr~$T_w$. 
This is not the case for $\delta = 1.2$ while the queue at $S$ is stable and $M_U \geq 400$. 
For every value of $w \in \{1/4, 2/4, 3/4 \}$, user $U$ should only be assisted by $S$ to achieve the \maxThr~$T_w$ since $q^*_C=1$ and $q^*_D = 0$. 
We also observe that, in this case, $S$ should more frequently assist $D$ since $q^*_S$ has almost always a higher value compared to $\delta=0.5$. 

\begin{table}[htbp]\caption{The values of $(q^*_S, q^*_C, q^*_D)$ that maximize the weighted sum throughput $T_w$ for different values of $M_U$ when $\alpha = 0.7$ and the queue at $S$ is stable.}
	\label{table:StableThr_vs Mu}
	\begin{center}
	\def\arraystretch{1.2}
	\begin{tabular}{ | c || c | c | c | c |  } 
			\hline \multicolumn{5}{|c|}{$\delta = 0.5$} \\ \hline
\multicolumn{1}{c}{$M_U $}  &  \multicolumn{1}{c}{} & \multicolumn{1}{c}{$ w = 1/4$} & \multicolumn{1}{c}{$ w = 2/4$} &  \multicolumn{1}{c}{$ w = 3/4$} \\ \hline
		
%		$ M_U$ &   & max. $T$ $(q^*_S, q^*_C, q^*_D)$ & max. $T$ $(q^*_S,  q^*_C, q^*_D)$   \\ \hline
		
		$ 200$ & max. $T_w$ & $0.227$ 	& $0.285$  		  & $0.342$ \\ 
		$ 		$  & $(q^*_S, q^*_C, q^*_D)$  & $(0.999, 1, 1)$	& $(0.999,1,1)$ & $( 0.999, 1, 1)$ \\ \hline
		
		$ 400$ & max. $T_w$ & $0.224$ 	   & $0.283$  		  & $0.341$ \\
		$ 		 $ & $(q^*_S, q^*_C, q^*_D)$  & $(0.801, 1, 1 )$  & $(0.824,1,1)$  & $( 0.769, 1,1)$ \\ \hline

		$ 600$ & max. $T_w$ & $0.221$ 	  	& $0.281$ 		   & $0.340 $ \\
		$   	 $ & $(q^*_S, q^*_C, q^*_D)$  & $(0.756, 1, 1)$  &  $(1, 1, 1)$  		 & $(1,1,1)$ \\ \hline
				
		$ 800$ & max. $T_w$ & $0.219$ 	   & $0.279$ 		  & $0.340$  \\
		$ 		 $ & $(q^*_S, q^*_C, q^*_D)$  & $(1, 1, 1)$ 		   & $(1, 1, 1)$  		  & $(0.746, 1, 1)$ \\ \hline
		
		$1000$ & max. $T_w$ & $0.217$ 	   & $0.278$ 		  & $0.339$ \\	
		$		  $ & $(q^*_S, q^*_C, q^*_D)$ & $(0.732, 1, 1)$  & $(1, 1, 1)$  		  & $(1, 1, 1)$ \\
	\end{tabular}  

	\begin{tabular}{ | c || c | c | c | c |  } \hline
	\hline \multicolumn{5}{|c|}{$\delta = 1.2$} \\ \hline
	\multicolumn{1}{c}{$M_U $}  &  \multicolumn{1}{c}{} & \multicolumn{1}{c}{$ w = 1/4$} & \multicolumn{1}{c}{$ w = 2/4$} &  \multicolumn{1}{c}{$ w = 3/4$} \\ \hline
	
		$ 200$  & max. $T_w$ 							 & $ 0.145$  	 	  & $ 0.230 $ 		 & $ 0.315 $ \\ 
		$  		$   & $(q^*_S, q^*_C, q^*_D)$ & $ (0.713, 1, 1)$ & $( 0.713, 1, 1)$  & $(0.713, 1, 1)$ \\ \hline
		
		$ 400$ & max. $T_w$ 							& $ 0.135$    	 	& $ 0.223 $ 			& $ 0.312$ \\
		$ $  	  & $(q^*_S, q^*_C, q^*_D)$	 & $(1, 1, 0)$  	  & $(0.999, 1, 0)$ & $(0.999, 1, 0)$ \\ \hline
		
		$ 600$ & max. $T_w$ 							& $ 0.129$ 		   	& $ 0.220 $   & $ 0.310 $ \\
		$ 		 $ & $(q^*_S, q^*_C, q^*_D)$  & $ (1, 1, 0)$    	 & $(1, 1, 0)$  & $(1, 1, 0)$ \\ \hline
		
		$ 800$ & max. $T_w$  							& $ 0.126$   	& $ 0.217 $ 	& $0.309$ \\
		$ 		 $ & $(q^*_S, q^*_C, q^*_D)$  & $(1,1,0)$     & $(1, 1, 0)$   & $(1,1,0)$ \\ \hline
		
		$1000$ & max. $T_w$  							& $0.123$    & $ 0.216 $   & $0.308$ \\	
		$ 		 $ & $(q^*_S, q^*_C, q^*_D)$  & $(1,1,0)$   & $(1, 1, 0)$   & $(0.540,1,0)$ \\ \hline		
	\end{tabular} 
\end{center}
\end{table}

In case the queue at $S$ is unstable, the values of $(q^*_S, q^*_C, q^*_D)$ for which the \maxThr~$T'_w$ is achieved can be found in Table \ref{table:UnstableThr_Mu}.
We observe that neither helper $S$ should serve helper $D$ ($q^*_S=0$) nor the latter should assist $U$ ($q^*_D=0$) to maximize $T'_w$ when (i) $\delta=0.5$ and $w=1/4$ for any cache  size $M_U$ or (ii) $\delta=0.5, w=2/4$ and user $U$'s cache can hold $M_U = 200$ files. 

Moreover, when $\delta = 0.5, M_U \geq 400$ and $w \in \{ 2/4, 3/4 \}$, helper $S$ should only serve helper $D$ and the latter should assist user $U$ since $(q^*_S, q^*_D) = (1,1)$. However, helper $S$ should slightly assist $U$ in some cases when e.g., $M_U = 400$ or $600$.
When $\delta=1.2$, helper $S$ should only serve the destination helper $D$ and the latter should assist user $U$ for any value of $M_U$ and $w$. 
Additionally, helper $S$ should not assist user $U$ for any cache size $M_U$ but $400$.

Furthermore, it should be noted that, for any value of $M_U$, the \maxThr~is decreasing as $\delta$ increases when $w=1/4$ and increases as $\delta$ increases when $w \in \{ 2/4, 3/4\}$.

\begin{table}[htbp]\caption{The values of $(q^*_S, q^*_C, q^*_D)$ that maximize the weighted sum throughput $T_w$ for different values of $M_U$ when $\alpha = 0.7$ and the queue at $S$ is unstable.}
	\label{table:UnstableThr_Mu}
	\begin{center}
		\def\arraystretch{1.2}
		\begin{tabular}{ | c || c | c | c | c |  } 
			\hline \multicolumn{5}{|c|}{$\delta = 0.5$} \\ \hline
			\multicolumn{1}{c}{$M_U $}  &  \multicolumn{1}{c}{} & \multicolumn{1}{c}{$ w = 1/4$} & \multicolumn{1}{c}{$ w = 2/4$} &  \multicolumn{1}{c}{$ w = 3/4$} \\ \hline
			$ 200 $ & max. $T'_w$  & $0.430$     &  $0.286$ &  $0.399$ \\
			$ $ 		& $(q^*_S, q^*_C, q^*_D)$  &  $(0,1,0)$    &  $(0,1,0)$ &  $(1,0.024,1 )$ \\ \hline
			$ 400 $ & max. $T'_w$  & $0.398$     &  $0.289$ &  $0.404$ \\
			$ $ 		& $(q^*_S, q^*_C, q^*_D)$  &  $(0,1,0)$    &  $(1,0.029,1)$ &  $(1,0.005,1)$ \\ \hline
			$ 600 $ & max. $T'_w$  & $0.374$     &  $0.295$ &  $0.410$ \\
			$ $ 		& $(q^*_S, q^*_C, q^*_D)$  &  $(0,1,0)$    &  $(1,0,1)$ &  $(1, 0.011, 1 )$ \\ \hline
			$ 800 $ & max. $T'_w$  & $0.354$     &  $0.295$ &  $0.416$ \\
			$ $ 		& $(q^*_S, q^*_C, q^*_D)$  &  $(0,1,0)$    &  $(1,0,1)$ &  $(1,0,1)$ \\ \hline
			$ 1000 $ & max. $T'_w$  & $0.337$     &  $0.298$ &  $0.422$ \\
			$ $ 		& $(q^*_S, q^*_C, q^*_D)$   &  $(0,1,0)$    &  $(1,0,1)$ &  $(1,0,1)$ \\
		\end{tabular}  		
		\begin{tabular}{ | c || c | c | c | c |  } \hline
			\hline \multicolumn{5}{|c|}{$\delta = 1.2$} \\ \hline
			\multicolumn{1}{c}{$M_U $}  &  \multicolumn{1}{c}{} & \multicolumn{1}{c}{$ w = 1/4$} & \multicolumn{1}{c}{$ w = 2/4$} &  \multicolumn{1}{c}{$ w = 3/4$} \\ \hline
			$ 200 $ & max. $T'_w$  & $0.189$ & $0.363$ & $0.537$ \\
			$ $ 		& $(q^*_S, q^*_C, q^*_D)$  & $(1,0,1)$ & $(1,0,1)$ & $(1,0,1)$ \\ \hline
			$ 400 $ & max. $T'_w$  & $0.190$ & $0.368$ & $0.546$  \\
			$ $ 		& $(q^*_S, q^*_C, q^*_D)$  & $(1,0.046,1)$ & $(1,0,1)$ & $(1,0.484,1)$  \\ \hline
			$ 600 $ & max. $T'_w$  & $0.190$ & $0.371$ & $0.552$  \\
			$ $ 		& $(q^*_S, q^*_C, q^*_D)$  & $(1,0,1)$ & $(1,0,1)$ & $(1,0,1)$  \\ \hline
			$ 800 $ & max. $T'_w$  & $0.191$ & $0.373$ & $0.556$ \\
			$ $ 		& $(q^*_S, q^*_C, q^*_D)$ & $(1,0,1)$ & $(1,0,1)$ & $(1,0,1)$  \\ \hline
			$ 1000 $& max. $T'_w$  & $0.191$ & $0.375$ & $0.559$  \\
			$ $ 		& $(q^*_S, q^*_C, q^*_D)$   & $(1,0,1)$ & $(1,0,1)$ & $(1,0,1)$  \\ \hline
		\end{tabular} 
		
	\end{center}
\end{table}

%\newpage
\subsection{Maximum Weighted Sum Throughput vs. Average Arrival Rate $\lambda$ when $M_D = 0$}
%{MAXIMUM WEIGHTED SUM THROUGHPUT vs. AVERAGE ARRIVAL RATE $\lambda$ when CACHE SIZE $M_D = 0$. 

We consider a scenario where the system parameters are the same as in Section \ref{sec:Throughput_vs_lambda} (see Tables \ref{table:links_params} and \ref{table:caches_params}), but helper $D$ cannot assist user $U$ since its cache cannot hold any files i.e., $M_D = 0$. Consequently, $\qd=0$ and $\phd=0$ as well.
This scenario will allow the study of the \maxThr~versus $\lambda$ when only one of the two helpers, the least powerful, is unable satisfy $U$'s needs for content from external resources.
\begin{figure}[ht] 
	\centering
	\subfloat[$\delta = 0.5$]{
		\includegraphics[width=1.05\linewidth]{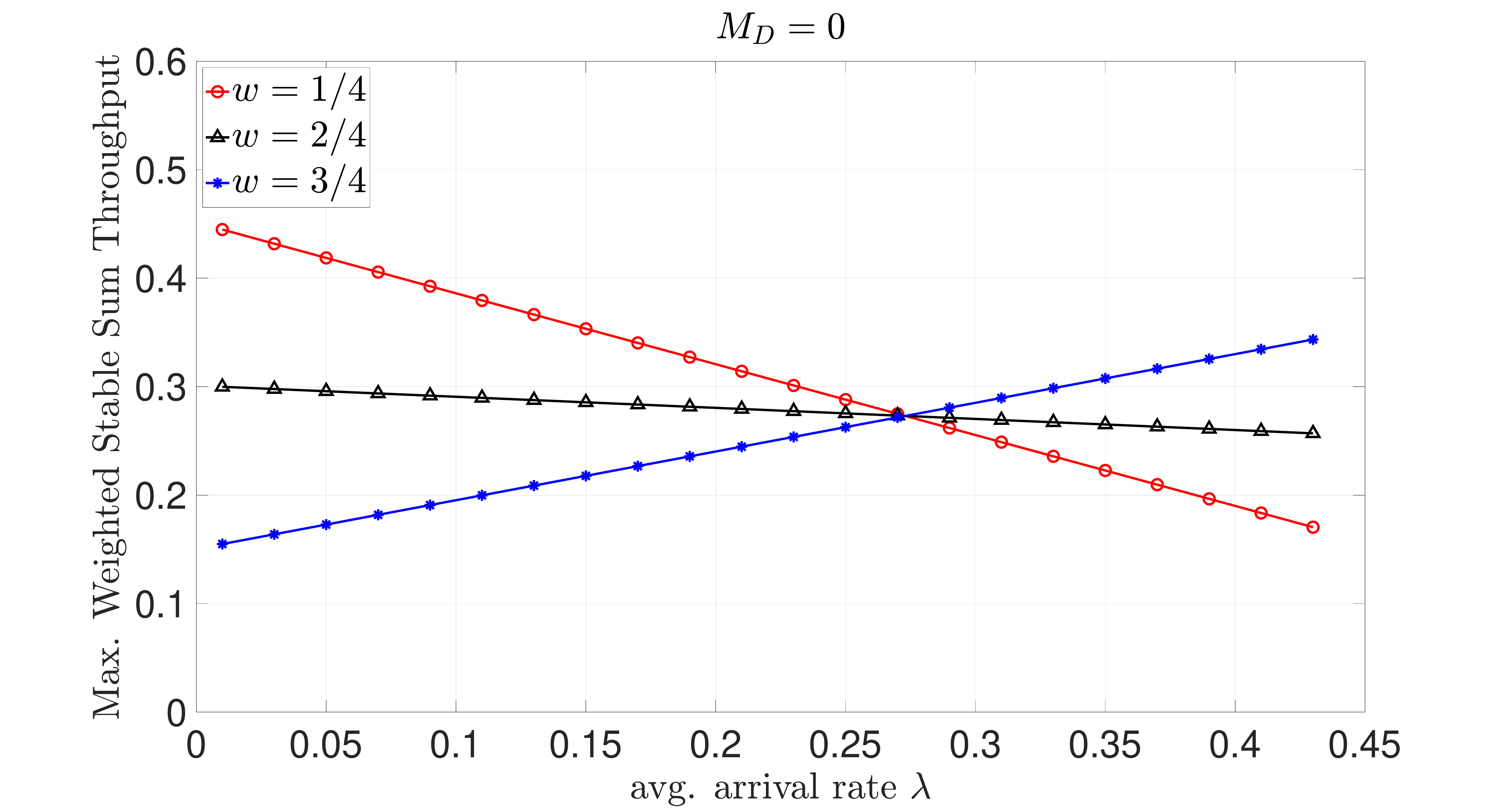}}
	\hfill
	\subfloat[$\delta = 1.2$]{
		\includegraphics[width=1.05\linewidth]{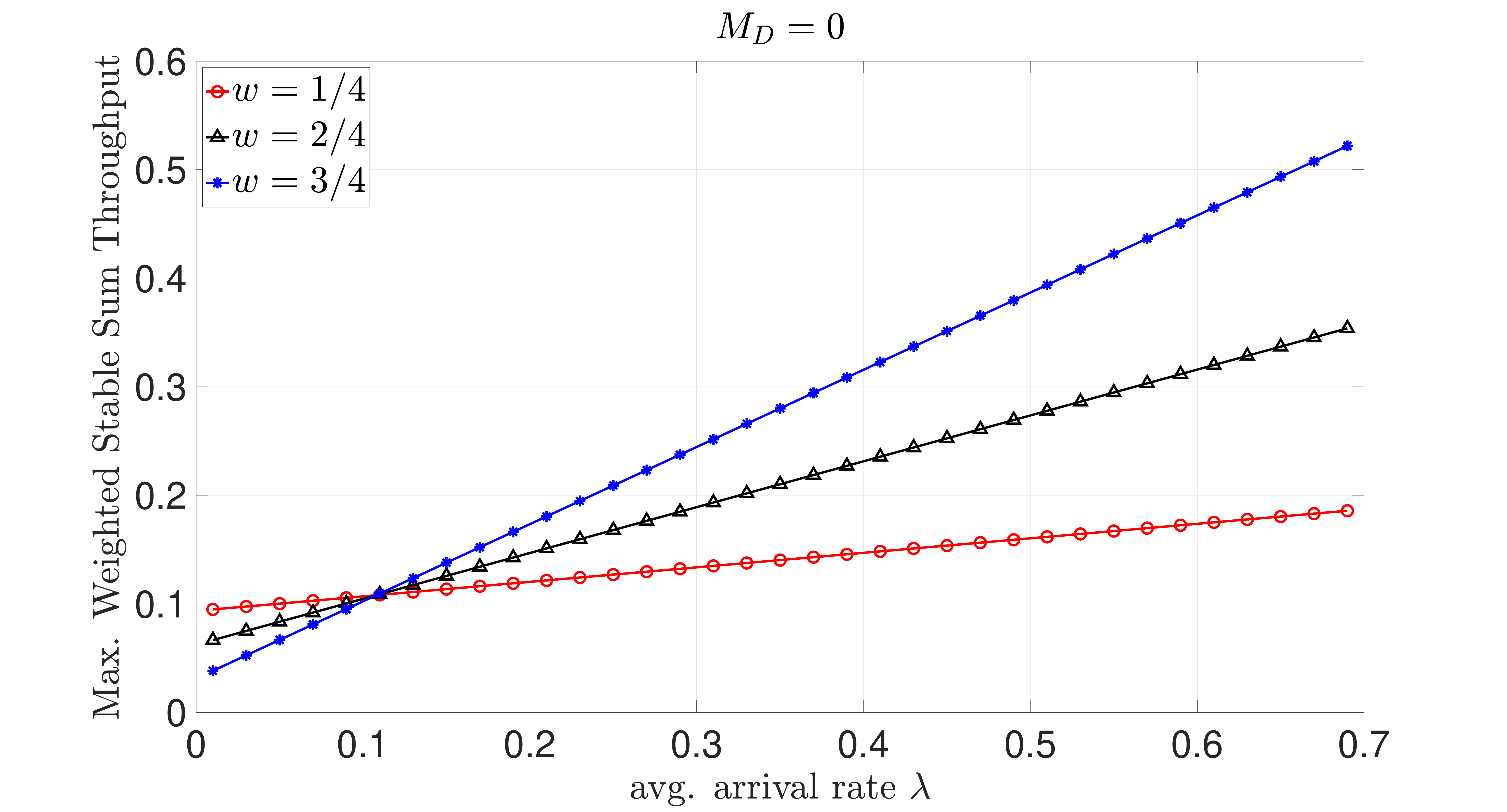}}	
	\caption{The \maxThr ~vs the average arrival rate $\lambda$ for which the queue at $S$ is stable, $ M_U = 200, M_D=0, M_S =2000,$ and $\alpha=0.7$ for: (a) $ \delta = 0.5$ and (b) $  \delta = 1.2 $.}
	\label{Fig:stableThr_vs_lambda_M_D_0}
\end{figure} 

In \figurename \ref{Fig:stableThr_vs_lambda_M_D_0}, we plot the \maxThr~versus $\lambda$ when the queue at $S$ is stable and $M_D=0$. We observe that, when $\delta=0.5$, the \maxThr~ (i) is a decreasing function of $\lambda$ for $w = 1/4$, (ii) slightly decreases for $w = 2/4$, and (iii) increases for $w = 3/4$. Recall that, by definition, in the first case $T_U$ dominates $T_S$, in the second case both thoughput terms contribute equally, and in the third case $T_S$ dominates $T_U$. 
When $\delta=1.2$, the \maxThr~is increasing with $\lambda$. We observe that higher values of $w$ yield steeper increases in the \maxThr.

When the queue at $S$ is stable, the \maxThr~is always achieved when $ q^*_C = 1 $ for any $w, \delta,$ and $\lambda$ using the system parameters we quoted before. 
However, in case $\delta =0.5$, helper $S$ should nearly always assist $U$ since $q^*_S \in \{0.977, 0.999\}$. When $\delta=1.2$, helper $S$ should always assist $U$ as Table \ref{table:stableThr_vs_lambda_MD0} depicts.  

On the other hand, when the queue at $S$ is unstable and $\delta=0.5$, helper $S$ should only assist $U$ when $w \in \{1/4, 2/4 \}$ and only assist $D$ when $w=3/4$ (see Table \ref{table:unstableThr_vs_lambda_MD0}). 
This is expected since in the latter case, $T_S$ dominates $T_U$ and, hence, it is preferable that $S$ always serves $D$ to maximize the contribution of $T_S$. 
In this case, if user $U$ requests content from external resources, it will be only served by the data center.
Moreover, when $\delta = 1.2$, it is optimal that helper $S$ serves only $D$ for any value of $w$.

\begin{table}[t!]\caption{The values of $q^*_S$ and $ q^*_C$ for which the weighted sum throughput is maximized when the queue at $S$ is stable, $\alpha = 0.7, M_U = 200, M_D = 0,$ and $ M_S = 2000$.}
	\def\arraystretch{1.1}%  1 is the default
	\begin{center}
	\begin{tabular}{ | c || c | c | c || c | c | c |  } \hline 
	$M_D =0 $&\multicolumn{3}{c||}{$\delta = 0.5$} & \multicolumn{3}{c|}{$\delta = 1.2$}  \\ \hline		
	\hline w & max.  $T_w$  &  $q^*_S$ & $q^*_C$ & max.  $T_w$  &  $q^*_S$ & $q^*_C$ \\ \hline
			$ 1/4 $  & $ 0.445 $ & $0.994$ 	& $1$ & $ 0.187 $  & $ 1 $ & $1$ \\ \hline
			$ 2/4 $ & $ 0.300$  & $0.977$	& $1$ & $ 0.358 $  & $ 1 $ & $1$ \\ \hline
			$ 3/4 $ & $ 0.348 $  & $0.999$ 	& $1$ & $ 0.529 $ &  $ 1 $ & $1$  \\ \hline
	\end{tabular}
	\end{center}

	\label{table:stableThr_vs_lambda_MD0}
\end{table}	

\begin{table}[ht] \caption{The values of $q^*_S$ and $q^*_C$ for which the weighted sum throughput is maximized when the queue at $S$ is unstable, $\alpha = 0.7, M_U = 200, M_D = 0,$ and $M_S = 2000$.}	
	\def\arraystretch{1.2} %  1 is the default
	\begin{center}
	\begin{tabular}{ | c || c | c | c || c | c | c |  } \hline 
	$M_D =0 $&\multicolumn{3}{c||}{$\delta = 0.5$} & \multicolumn{3}{c|}{$\delta = 1.2$}  \\ \hline
		\hline w  & max. $T'_w$ &  $q^*_S$ & $q^*_C$ & max. $T'_w$ &  $q^*_S$ & $q^*_C$  \\ \hline 
		$ 1/4 $ & $ 0.451 $ & $0$ & $1$ & $0.187$  & $1$ & $0$\\ \hline
		$ 2/4 $ & $ 0.301 $ & $0$ & $1$ & $0.359$  & $1$ & $0$ \\ \hline
		$ 3/4 $ & $ 0.349 $ & $1$ & $0$ & $0.531$  & $1$ & $0$\\ \hline
	\end{tabular}
	\end{center}
	\label{table:unstableThr_vs_lambda_MD0}
\end{table}

\subsection{Maximum Weighted Sum Throughput vs. Average Arrival Rate $\lambda$ when $M_s = 0$}
%{MAXIMUM WEIGHTED SUM THROUGHPUT vs. AVERAGE ARRIVAL RATE $\lambda$ when $M_s = 0$}
Here, we study the \maxThr~versus the average arrival rate $\lambda$ when node $S$ is not equipped with cache i.e., $M_S=0$, and, hence, $\qc=0$ and $\phs=0$.
The parameters of helper $D$'s cache and the wireless links can be found in Tables \ref{table:caches_params} and \ref{table:links_params}, respectively. 

\begin{figure}[htbp] 
	\centering
	\subfloat[$\delta = 0.5$]{
	\includegraphics[width=1.05\linewidth]{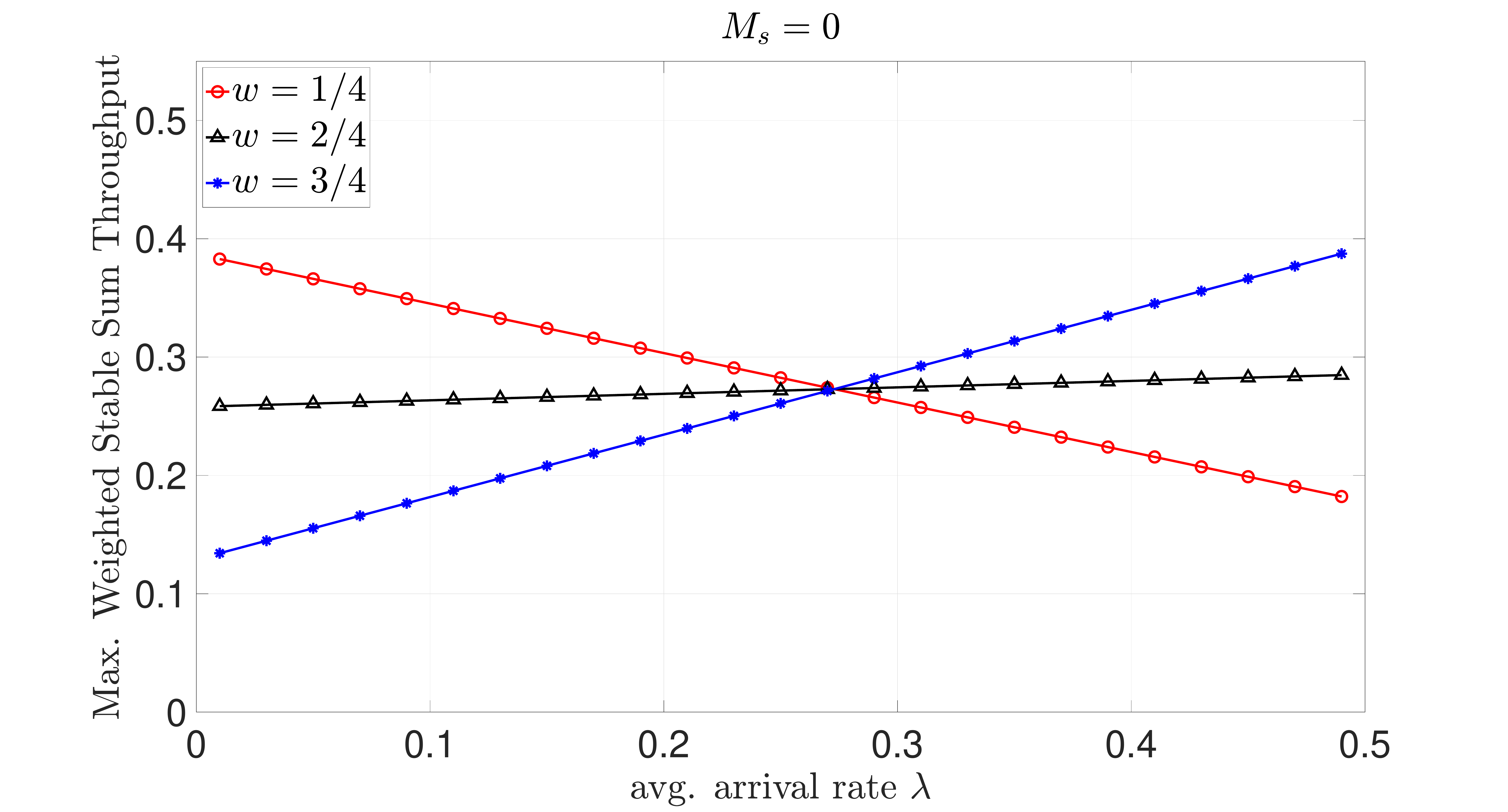}}
	\hfill
	\subfloat[$\delta = 1.2$]{
	\includegraphics[width=1.05\linewidth]{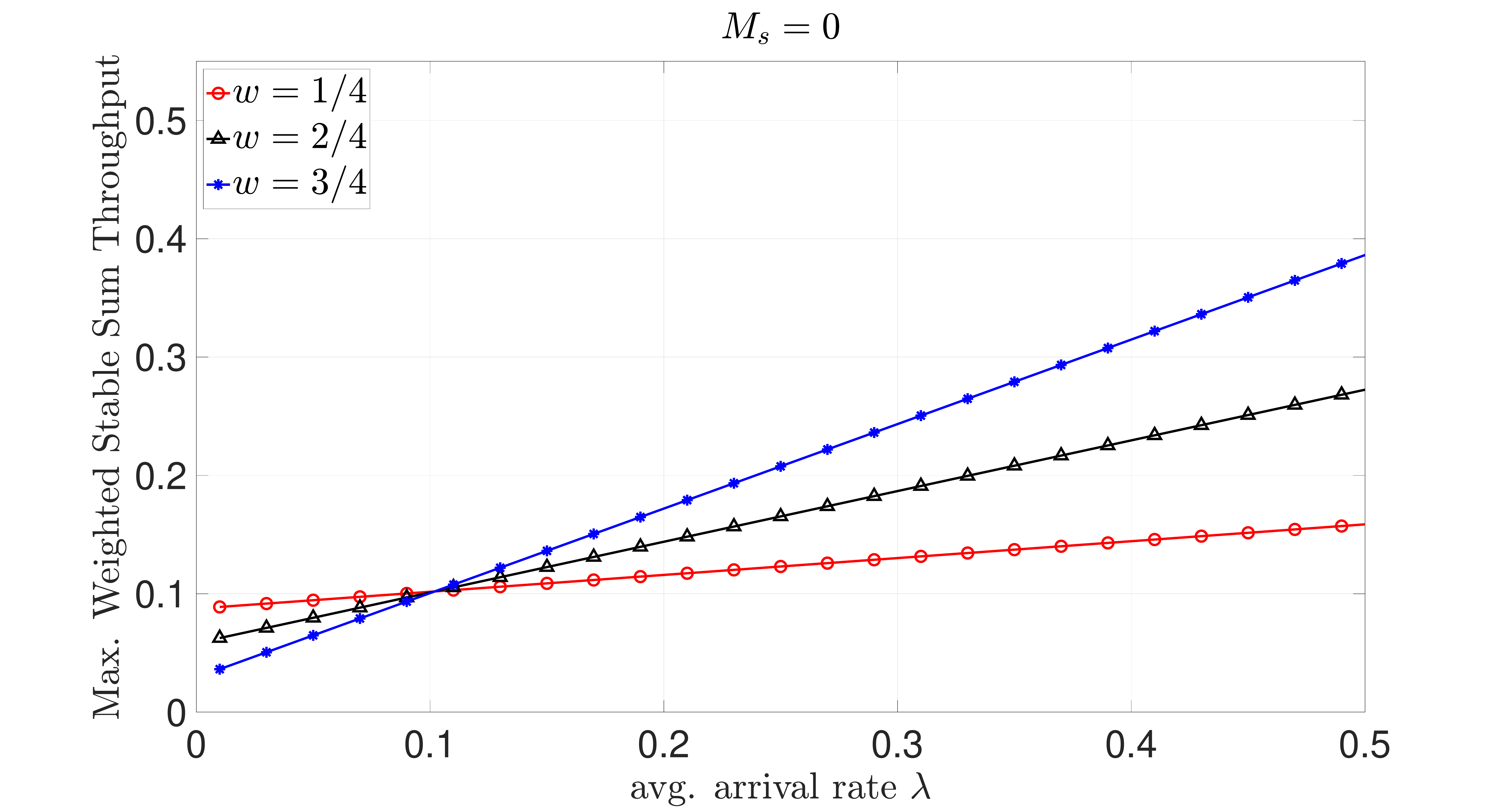}} 		
	\caption{The \maxThr ~vs. the average arrival rate $\lambda$ for which the queue at $S$ is stable, $M_U =200, M_D = 1000, M_S=0$ and $\alpha=0.7$ using different values of $w$ for: (a) $ \delta = 0.5$ and (b) $  \delta = 1.2 $.}
	\label{Fig:stableThr_vs_lambdaMS_0}
\end{figure} 

In \figurename~\ref{Fig:stableThr_vs_lambdaMS_0}, we plot the \maxThr~versus $\lambda$ for which the queue at $S$ is stable when $M_S=0$ for different values of $w$. 
Regarding $\delta = 0.5$, when $w=1/4$, the \maxThr~is decreasing with $\lambda$.
When $w \in \{2/4, 3/4 \} $, and $\delta \in \{0.5, 1.2\}$ or $w=1/4$ and $\delta=1.2$, the \maxThr~is an increasing function of $\lambda$.
%
%In case $w=1/4$ and $\delta=0.5$, the \maxThr is lower compared to the case that both helpers assist user $U$ (see Fig. \ref{Fig:stableThroughput_vs_lambda}) or when only helper $D$ assists $U$ requests for external content (see Fig. \ref{Fig:stableThr_vs_lambda_M_D_0}). However, when $S$ is not assisting $U$, the \maxThr~is lower compared to the other two cases.

\begin{table}[ht]	\caption{The values of $q^*_S$ and $ q^*_D$ for which the weighted sum throughput is maximized when the queue at $S$ is stable, $\alpha = 0.7, M_U = 200, M_D=1000,$ and $M_S = 0$.}
	\def\arraystretch{1.1}%  1 is the default
	\begin{center}
		\begin{tabular}{ | c || c | c | c || c | c | c |} \hline
			$M_S =0 $&\multicolumn{3}{c||}{$\delta = 0.5$} & \multicolumn{3}{c|}{$\delta = 1.2$}  \\ \hline 	
%			\hline  \multicolumn{4}{|c|}{$\delta = 0.5$ and $M_S = 0$} \\ \hline			
%			\hline  & max weighted sum  & &  \\
			\hline w & max.$T_w$  &  $q^*_S$ & $q^*_D$ & max.$T_w$  &  $q^*_S$ & $q^*_D$ \\ \hline
			$ 1/4 $  & $ 0.383$  & $ 0.993 $ 	& $1$ & $ 0.189 $  & $ 1 $ & $1$ \\ \hline
			$ 2/4 $ & $ 0.296$  &        $ 1  $	   & $1$ & $ 0.362 $  & $ 1 $ & $1$  \\ \hline
			$ 3/4 $ & $ 0.498$  & 		 $ 1  $    & $1$ & $ 0.536 $  & $ 1 $ & $1$ \\ \hline
%		\end{tabular}
%		\begin{tabular}{ | c || c || c | c | }
%			\hline  \multicolumn{4}{|c|}{$\delta = 1.2$ and $M_S= 0$} \\ \hline	
%			\hline  & max weighted sum  & &  \\
%			w & max.$T_w$  &  $q^*_S$ & $q^*_D$\\ \hline
%			\hline
%			$ 1/4 $  & $ 0.189 $  & $ 1 $ & $1$ \\ \hline
%			$ 2/4 $ & $ 0.362 $  & $ 1 $ & $1$ \\ \hline
%			$ 3/4 $ & $ 0.536 $  & $ 1 $ & $1$ \\ \hline
		\end{tabular}
	\end{center}
	\label{table:stableThr_vs_lambdaMS_0}
\end{table}

In Table \ref{table:stableThr_vs_lambdaMS_0}, we present the values of $q^*_S$ and $q^*_D$ that achieve \maxThr~when the queue at $S$ is stable for different values of $w$. Recall that, in this specific scenario, $q^*_C=0$ since helper $S$ has no cache and, thus, it cannot assist $U$. Therefore, $S$ is only useful to helper $D$.
We observe that the \maxThr~is lowered compared to the case when $M_D=0$ for $\delta = 0.5$ and slightly higher for $\delta=1.2$ (compare with Table \ref{table:stableThr_vs_lambda_MD0}). Additionally, helper $S$ should almost always serve  $D$ and the later should always assist user $U$ to achieve the \maxThr.
\begin{table}[ht] 	\caption{The values of $q^*_S$ and $q^*_D$ for which the weighted sum throughput is maximized when the queue at $S$ is unstable, $\alpha = 0.7, M_U = 200, M_D = 1000,$ and $M_S = 0$.}
	\def\arraystretch{1.1} %  1 is the default
	\begin{center}
		\begin{tabular}{ | c || c | c | c || c | c | c |  } \hline 
			$M_S =0 $&\multicolumn{3}{c||}{$\delta = 0.5$} & \multicolumn{3}{c|}{$\delta = 1.2$}  \\ \hline
			\hline w  & max.$T'_w$ &  $q^*_S$ & $q^*_D$ & max.$T'_w$ &  $q^*_S$ & $q^*_D$ \\ \hline 
			$ 1/4 $   & $ 0.387 $    & $0$ 		  &	$1$  		& $ 0.189$     & $1$ 		 & $1$  \\ \hline
			$ 2/4 $  & $ 0.286$ 	& $1$ 		  & $1$    		& $ 0.363$    & $1$ 		& $1$ \\ \hline
			$ 3/4 $  & $ 0.399$ 	& $1$ 		  & $1$ 	   & $ 0.537$  	  & $1$ 		& $1$ \\ \hline
	\end{tabular}
%	\begin{tabular}{ | c || c | c | c | }	\hline 
%			\multicolumn{4}{|c|}{$\delta = 1.2$ and $M_S = 0$} \\ \hline 
%			\hline w  & max.$T'_w$  &  $q^*_S$ & $q^*_D$ \\ \hline 
%			$ 1/4 $   & $ 0.189$  & $1$ & $1$ \\ \hline
%			$ 2/4 $  & $ 0.363$  & $1$ & $1$ \\ \hline
%			$ 3/4 $  & $ 0.537$  & $1$ & $1$  \\ \hline 
%		\end{tabular}
	\end{center}
	\label{table:unstableThr_vs_lambdaMS_0}
\end{table}	
%\begin{table}[ht]	
%	\def\arraystretch{1.1} %  1 is the default
%	\begin{center}
%		\begin{tabular}{ | c || c | c | c || c | c | c |  } \hline 
%			$M_S =0 $&\multicolumn{3}{|c|}{$\delta = 0.5$} & \multicolumn{3}{|c|}{$\delta = 1.2$}  \\ \hline
%			\hline w  & max.$T'_w$ &  $q^*_S$ & $q^*_D$ \\ \hline 
%			$ 1/4 $  & $ 0.387 $ & $0$ & $1$ \\ \hline
%			$ 2/4 $ & $ 0.286$ & $1$ & $1$ \\ \hline
%			$ 3/4 $ & $ 0.399$ & $1$ & $1$ \\ \hline
%		\end{tabular}
%			\begin{tabular}{ | c || c | c | c | }	\hline 
%					\multicolumn{4}{|c|}{$\delta = 1.2$ and $M_S = 0$} \\ \hline 
%					\hline w  & max.$T'_w$  &  $q^*_S$ & $q^*_D$ \\ \hline 
%					$ 1/4 $   & $ 0.189$  & $1$ & $1$ \\ \hline
%					$ 2/4 $  & $ 0.363$  & $1$ & $1$ \\ \hline
%					$ 3/4 $  & $ 0.537$  & $1$ & $1$  \\ \hline 
%				\end{tabular}
%	\end{center}
%	\caption{The values of $q^*_S$ and $q^*_D$ for which the weighted sum throughput is maximized when the queue at $S$ is \underline{unstable}, $\alpha = 0.7, M_U = 200, M_D = 1000,$ and $M_S = 0$.}
%	\label{table:unstableThr_vs_lambdaMS_0}
%\end{table}	
In Table \ref{table:unstableThr_vs_lambdaMS_0}, we present the values of $q^*_S$ and $q^*_D$ that achieve \maxThr~when the queue at $S$ is unstable for different values of $w$. 
In order to maximize the weighted sum throughput, helper $S$ should always serve $D$ for any values of $\delta$ and $w$ apart from the case in which $w=1/4$ and $\delta =0.5$ for which $S$ should remain silent since $q^*_S=0$. 
Furthermore, helper $D$ should always assist $U$ requests for every value of $w$ and $\delta$ we used.
The \maxThr~is higher compared to the case in which $M_D = 0$ (compare with Table \ref{table:unstableThr_vs_lambda_MD0}) for every value of $w$ and $\delta$ apart from the cases in which $\delta=0.5$ and $w \in \{ 1/4, 2/4 \}$.

\subsection{Average delay at user $U$}

\begin{figure}[t] 
		\centering
		\includegraphics[width=1.1\linewidth]{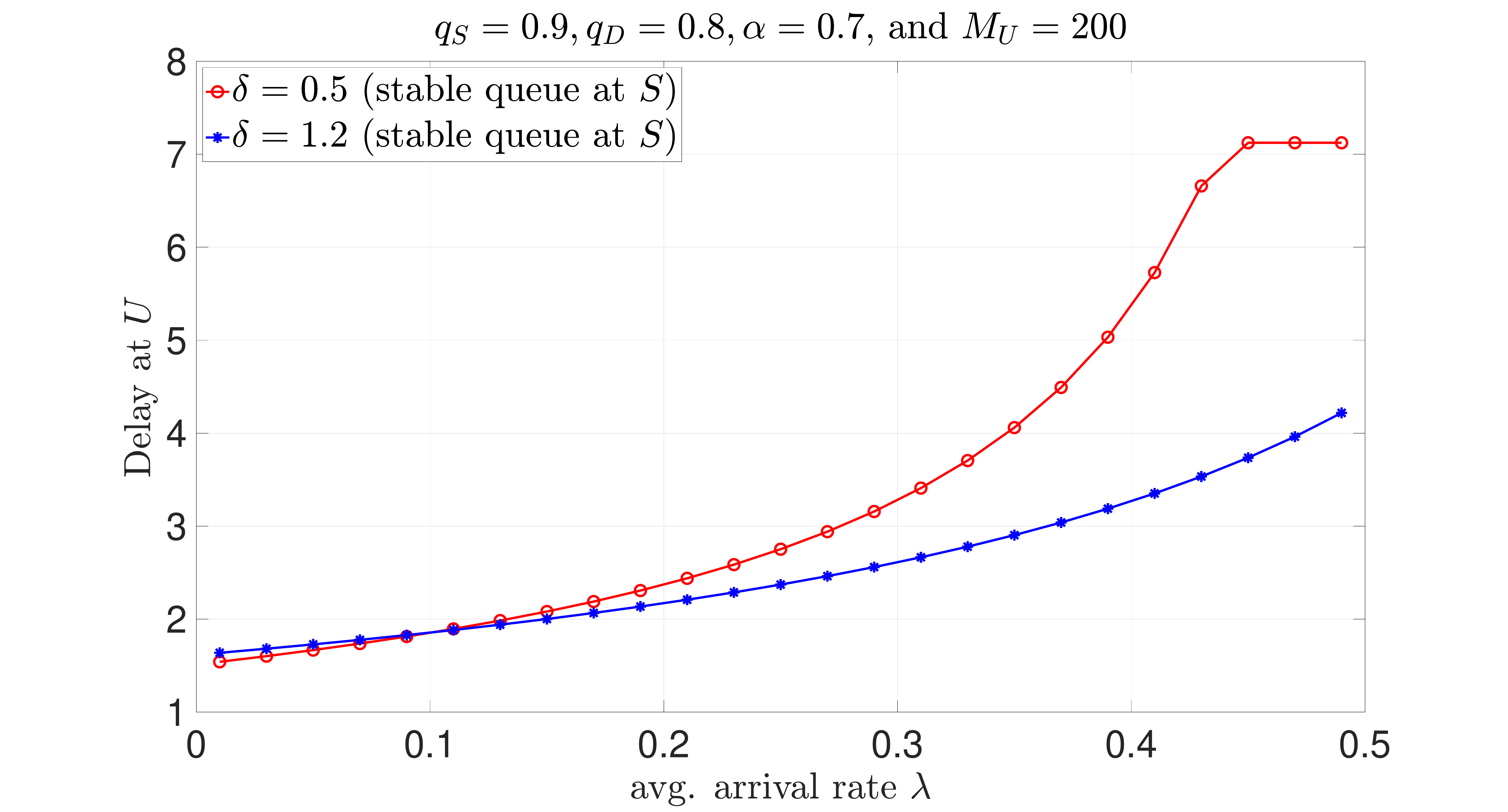}	
	\caption{The average delay at $U$ vs. average arrival rate $\lambda$ at $S$ for $\delta \in \{0.5, 1.2\}$. }
	\label{Fig:delay_vs_lambda}
\end{figure} 

Here, we present the numerical results of the average delay experienced by user $U$ to receive content from external sources. 
The delay analysis can be found in Section \ref{Sec:Delay}.

In the following plots, we study how the average arrival rate $\lambda$, the data center's random availability $\alpha$, the probability that $S$ attempts transmissions to $D$, $\qs$, the probability that $D$ attempts transmissions to $U$, $\qd$, and the cache size at $U$, $M_U$ affect the average delay at $U$.
The wireless links characteristics can be found in Table \ref{table:links_params}.
The cache sizes were set to hold $M_S = 2000$ and $M_D = 1000$ files at $S$ and $D$, respectively and we used two different values for $\delta$ to examine its effect on the realized average delay. Hence, the values of $\qu, \phd,$ and $\phs$ were given by (\ref{eq:qu}) - (\ref{eq:phs}) depending on $\delta$. Also, we set $q_c=0.5$.

In \figurename~\ref{Fig:delay_vs_lambda}, the average delay versus the arrival rate at helper $S$ is depicted for  $\qs = 0.9, \qd = 0.8, \alpha = 0.7$ and $M_U = 200$.
We observe that the delay increases with the arrival rate and the increase rate is steeper when $\delta=0.5$ compared to $\delta=1.2$. 
As we explained in Section \ref{Sec:CachePolicy}, higher values of $\delta$ yield more requests for a few most popular files. 
Therefore, for a given $M_U$, the higher the $\delta$, the lower the $\qu$ i.e., user $U$ requests files from external sources with lower probability, as well as lower value for cache hits $\phd$ and $\phs$ (for given $M_D$ and $M_S$). 
Fewer requests for files from external resources require fewer transmissions to $U$ and, hence, less interference is realized. Consequently, less average delay is experienced at $U$.

\begin{figure}[t]
	\centering
	\subfloat[$\lambda = 0.2$]{
	\includegraphics[width=1.05\linewidth]{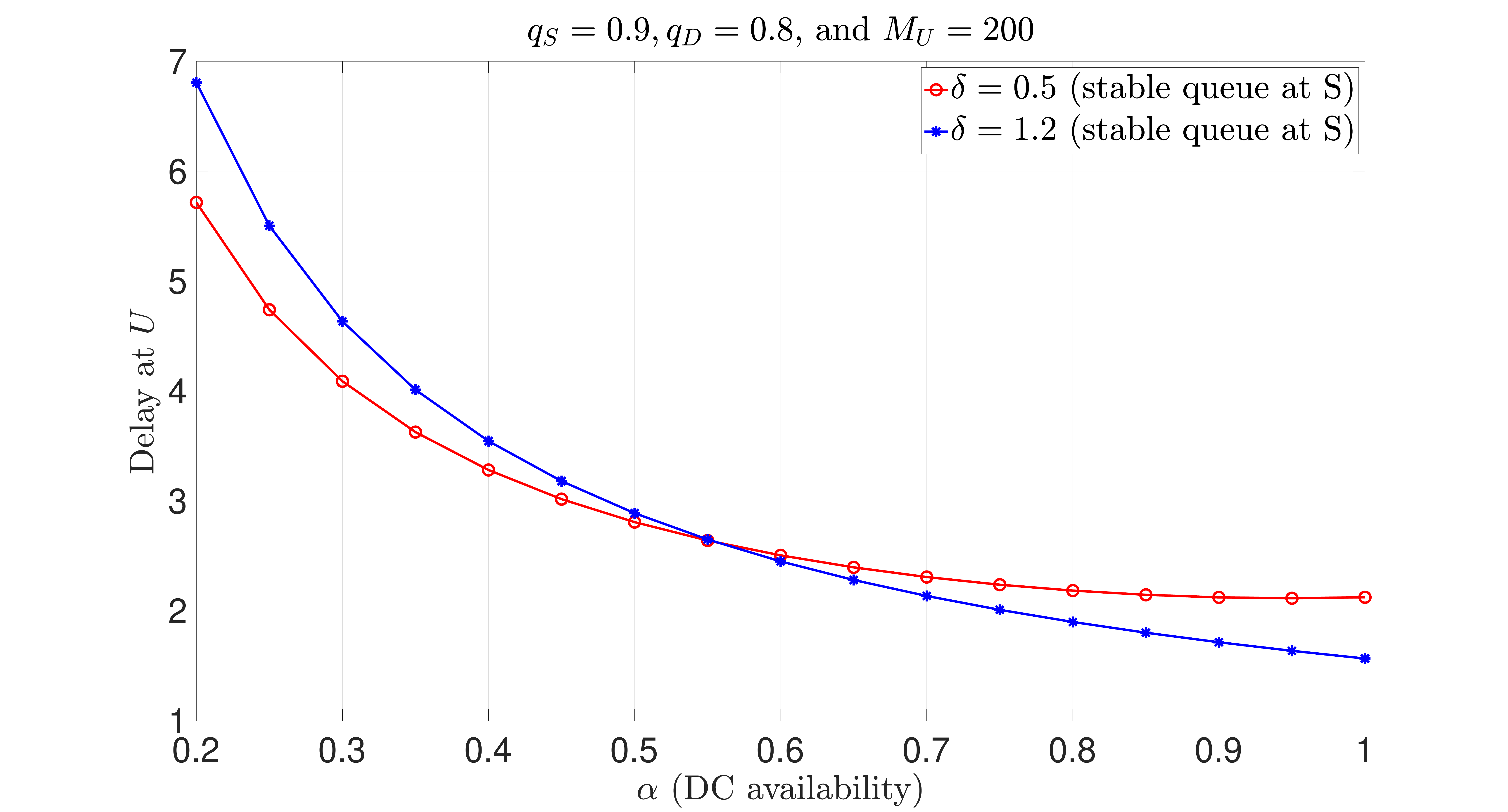}}
	\hfill
	 \subfloat[$\lambda = 0.4$]{
	 \includegraphics[width=1.05\linewidth]{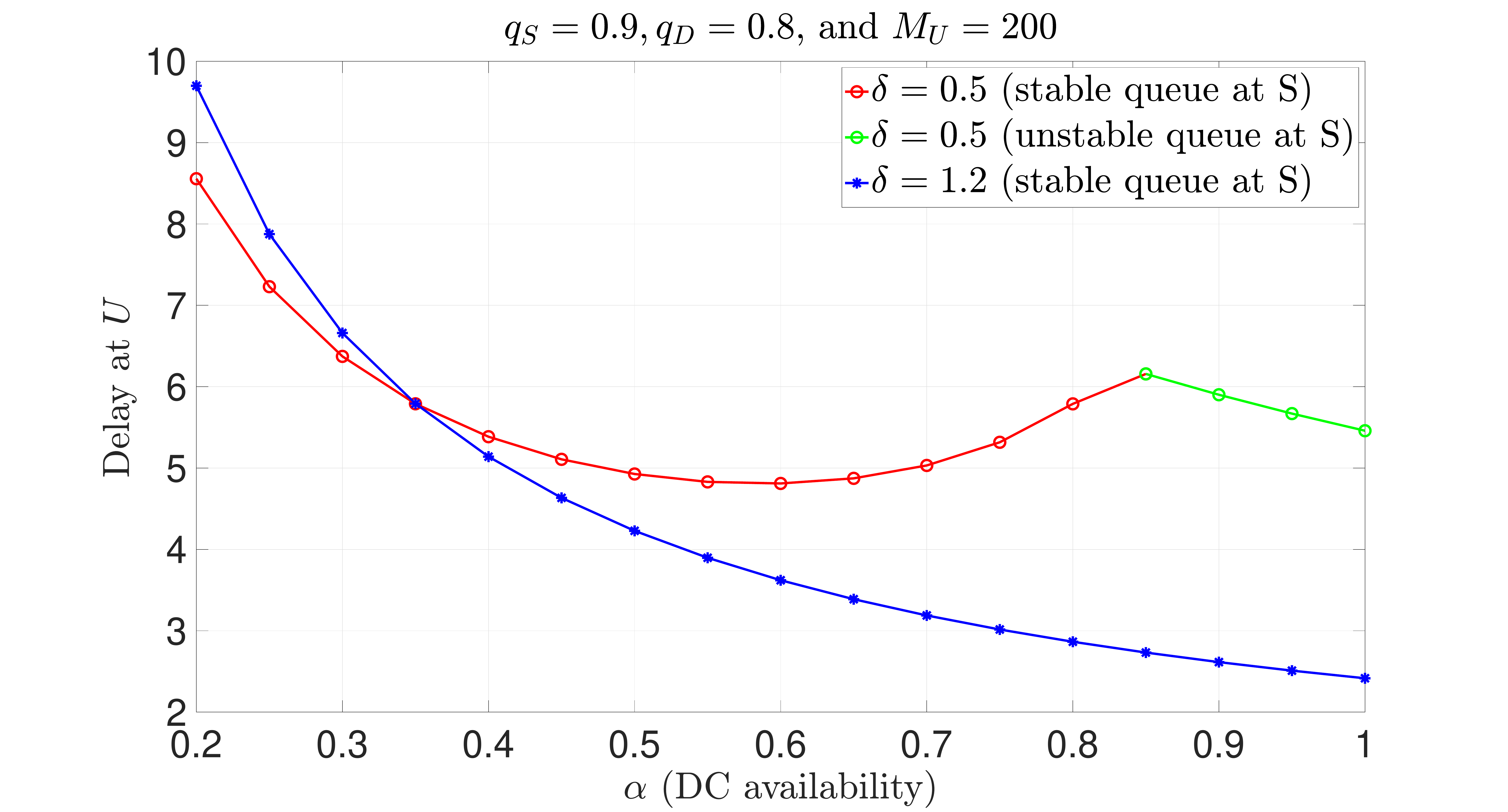}}
	\caption{The average delay at $U$ vs. data center's random availability $\alpha$ for $\delta \in \{0.5, 1.2\}$. }
	\label{Fig:delay_vs_alpha}
\end{figure}

In \figurename~\ref{Fig:delay_vs_alpha}, we present the average delay at $U$ versus data center's availability for two cases of arrival rate $\lambda = 0.2$ and $\lambda=0.4$.
We observe that the delay is lower when $\lambda = 0.2$ since a higher average arrival rate is more likely to create a congested queue at $S$ and, consequently, a higher delay. 
In case $\lambda = 0.2$, the delay is decreased with the increase of $\alpha$ and the queue at helper $S$ is stable for any $\alpha \in [0.2,1 ]$. Additionally, the decrease is steeper with $\alpha$ when $\delta=1.2$. 
When $\lambda = 0.4$ and $\delta=0.5$, the queue at $S$ remains stable for $\alpha  \in [0.2, 0.8]$ and the delay has the non-monotonic behavior of \figurename \ref{Fig:delay_vs_alpha}
(b). For $\alpha \in [ 0.8, 1]$, the average delay starts decreasing  with $\alpha$ and the queue at $S$ is unstable. When $\delta=1.2$, the queue at $S$ is stable for every value of $\alpha$ and the delay is decreased with the increased availability of the data center. 

\begin{figure}[t]
	\centering
	\subfloat[$\lambda = 0.2$]{	\includegraphics[width=1.05\linewidth]{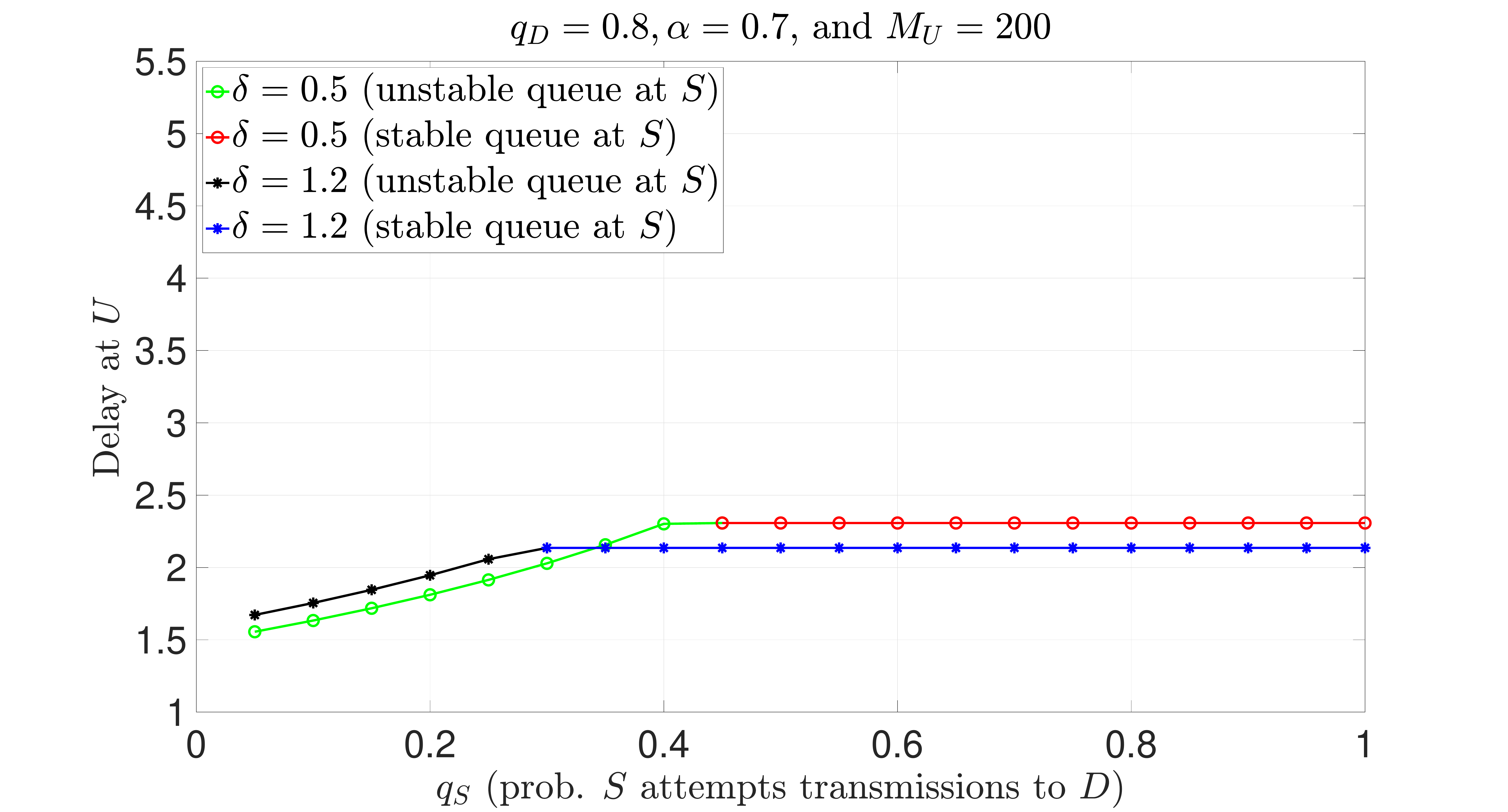}}
	\hfill
	\subfloat[$\lambda = 0.4$]{ \includegraphics[width=1.05\linewidth]{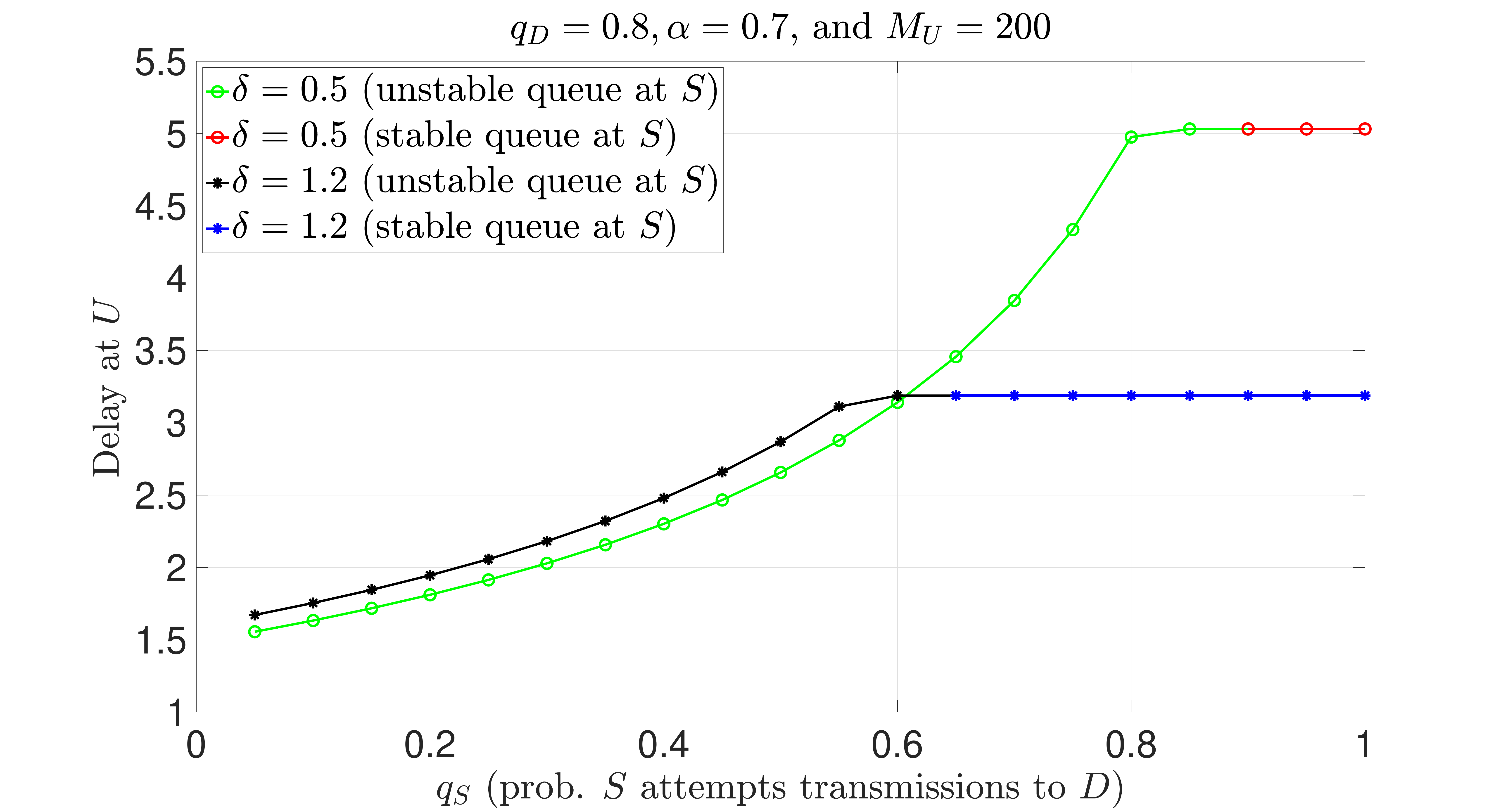}}
	\caption{The average delay at $U$ vs. $\qs$ for $\delta \in \{0.5, 1.2\}$. }
	\label{Fig:delay_vs_qs}
\end{figure} 

In \figurename~\ref{Fig:delay_vs_qs}, we plot the average delay at $U$ versus $\qs$ for $\lambda=0.2$ and $\lambda=0.4$. We observe that as long as the queue at $S$ is unstable, the delay increases with the $\qs$ increase. This is expected since as $\qs$ increases, helper $S$ attempts more transmissions to helper $D$ and, consequently, it is not only less likely to assist $U$ but also $U$'s probability to find an available helper is decreased (since the $S-D$ pair communicates). 
Regarding the case in which the queue at $S$ is stable, increasing $\qs$ does not contribute to delay's improvement. Moreover, a lower value of $\qs$ is required to achieve queue stability at $S$ when $\lambda=0.2$ compared to $\lambda=0.4$. This is expected since a higher average arrival rate requires a higher average service rate to maintain queue stability.

\begin{figure}[t]
	\centering
	\subfloat[$\lambda = 0.2$]{ \includegraphics[width=1.05\linewidth]{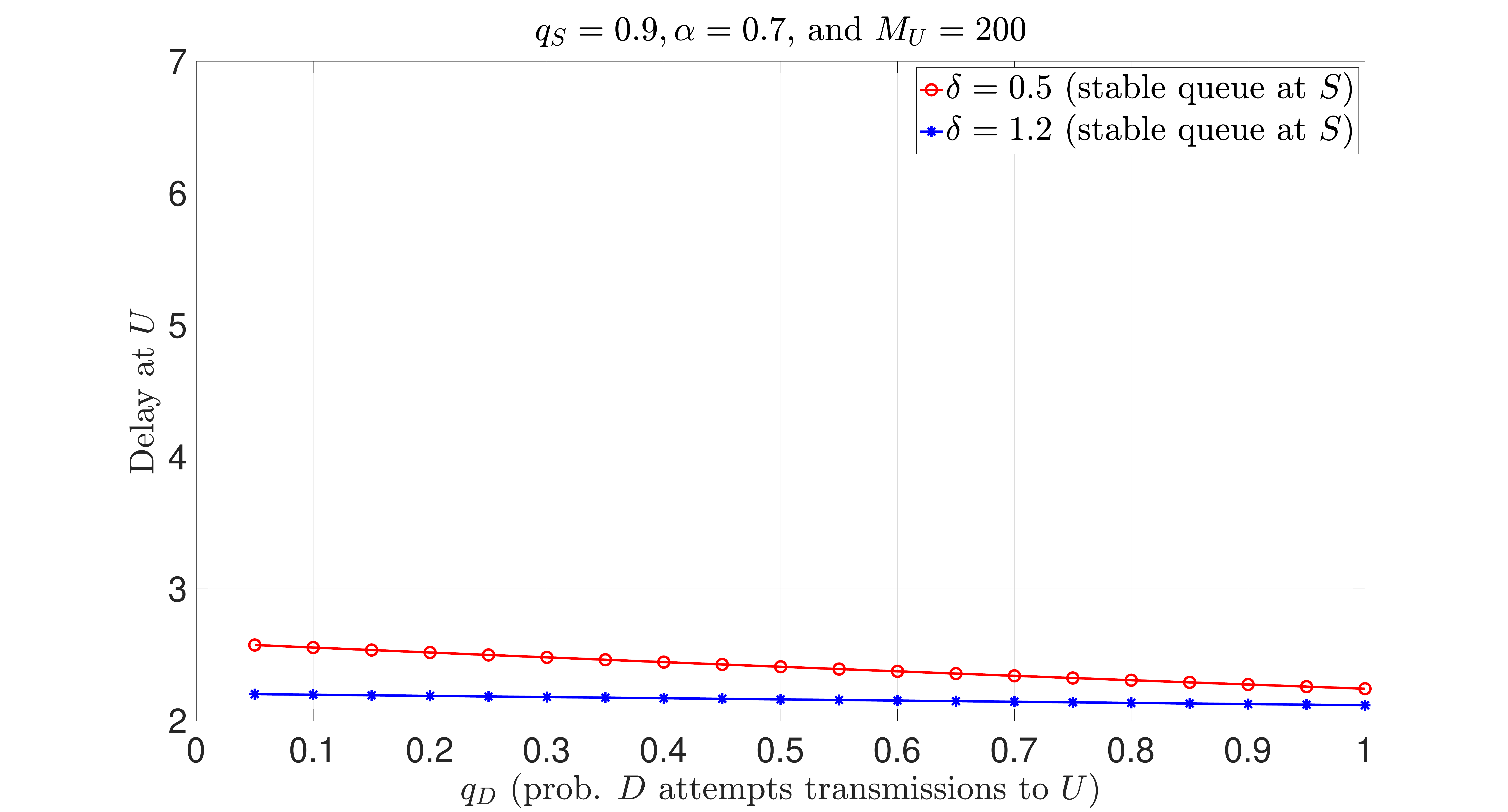}}
	\hfill
	\subfloat[$\lambda = 0.4$]{ \includegraphics[width=1.05\linewidth]{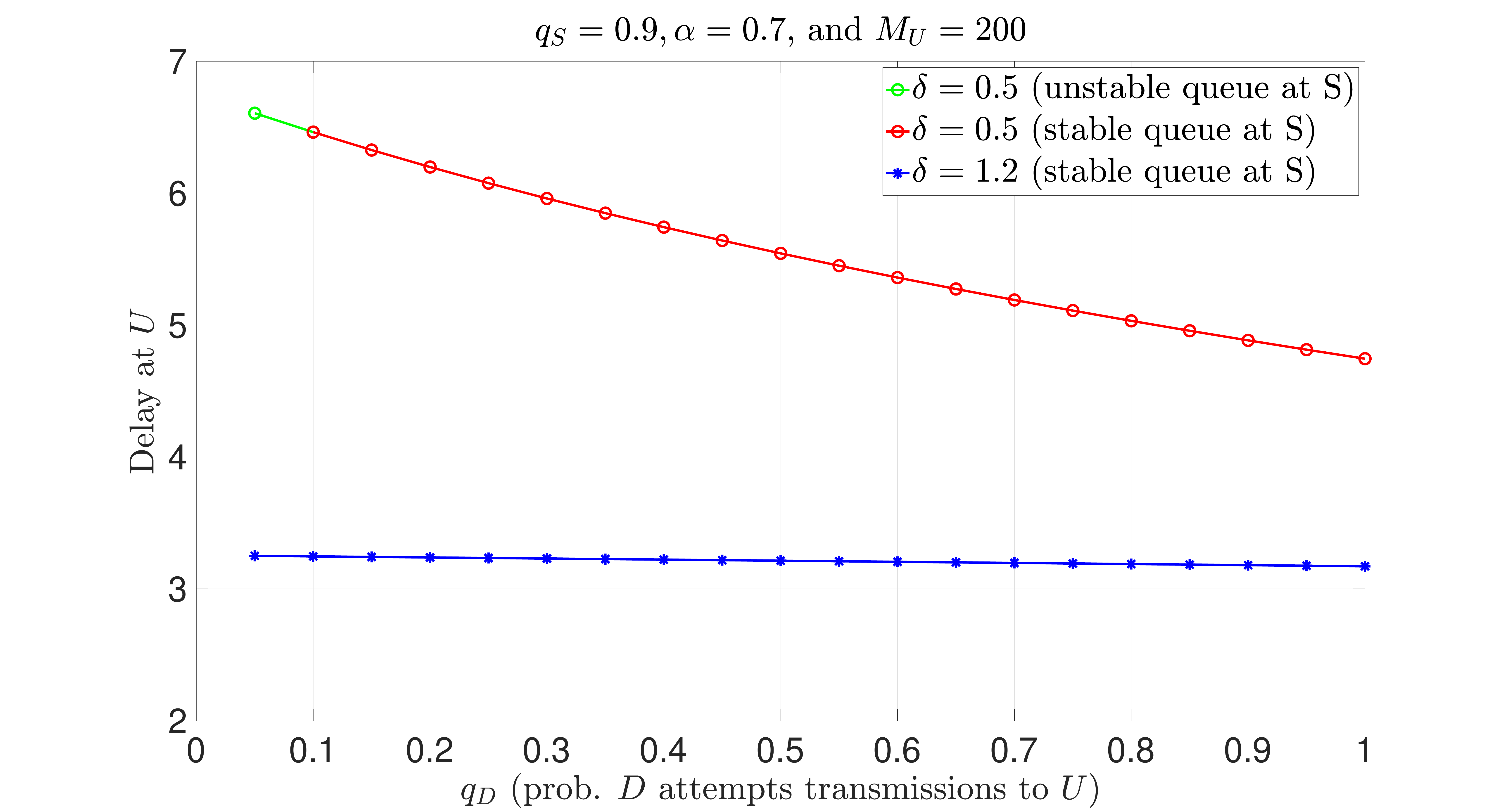}}
	\caption{The average delay at $U$ vs. $\qd$ for $\delta \in \{0.5, 1.2\}$. }
	\label{Fig:delay_vs_qd}
\end{figure} 

In \figurename~\ref{Fig:delay_vs_qd}, we demonstrate the average delay at $U$ versus $\qd$ for $\lambda=0.2$ and $\lambda=0.4$.
In the former case, the delay is slightly decreased with the increase of $\qd$
This can be attributed to helper $D$'s increased assistance that yields more transmissions to $U$ and, hence, potentially decreased delay.
When $\lambda=0.4$, the average delay decreases considerably with $\qd$ when $\delta = 0.5$ due to the increased assistance of helper $D$, but decreases slightly in case $\delta = 1.2$. 
This is expected, as we previously explained, since higher values of $\delta$ create more requests for a few most popular files, and, thus, $U$'s request for external content is decreased. As a result, the average delay at $U$ is decreased compared to lower $\delta$ values. 

\begin{figure}[t]
	\centering
	\subfloat[$\lambda = 0.2$]{	\includegraphics[width=1.05\linewidth]{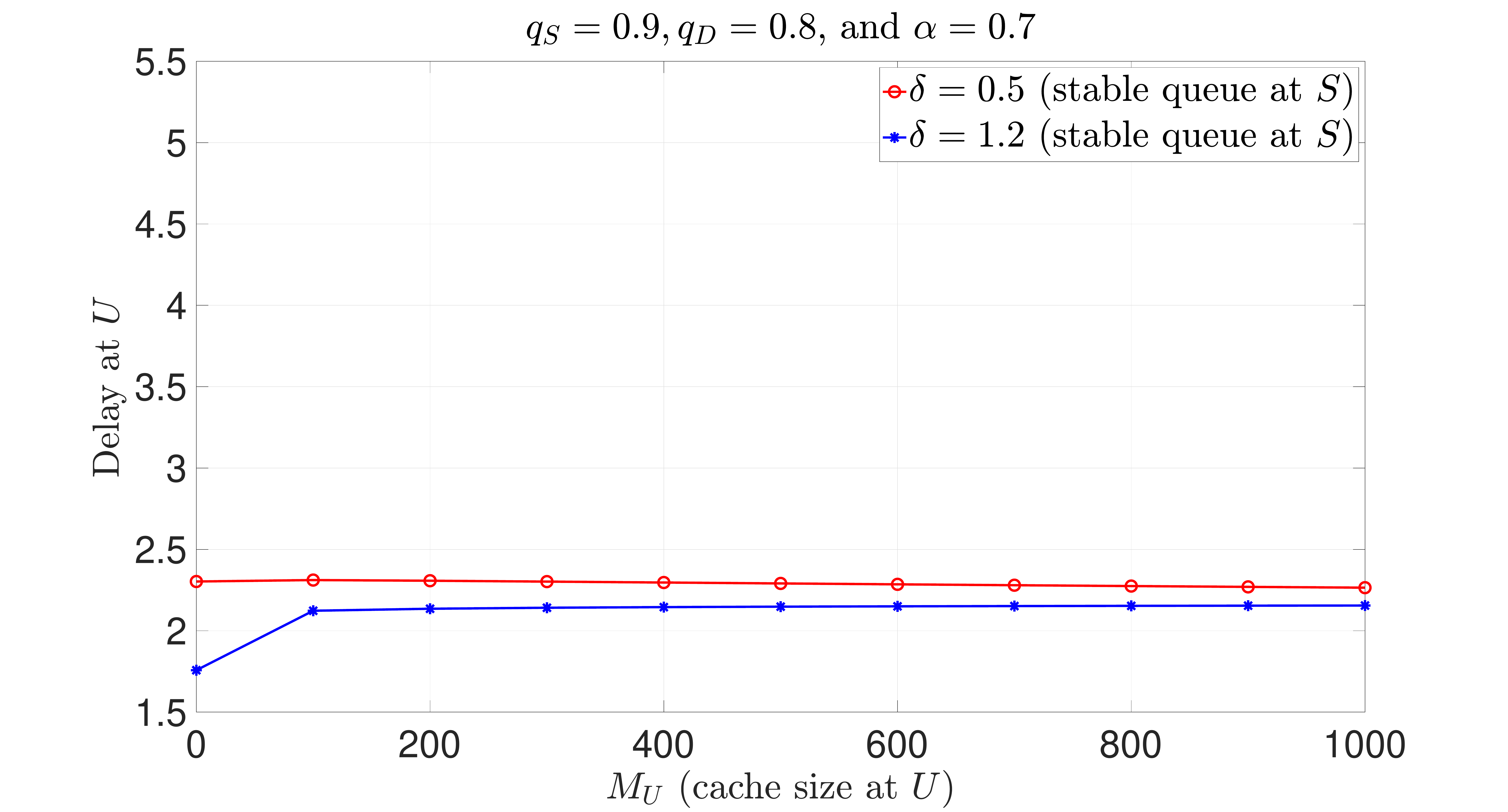}}
	\hfill
	\subfloat[$\lambda = 0.4$]{	\includegraphics[width=1.05\linewidth]{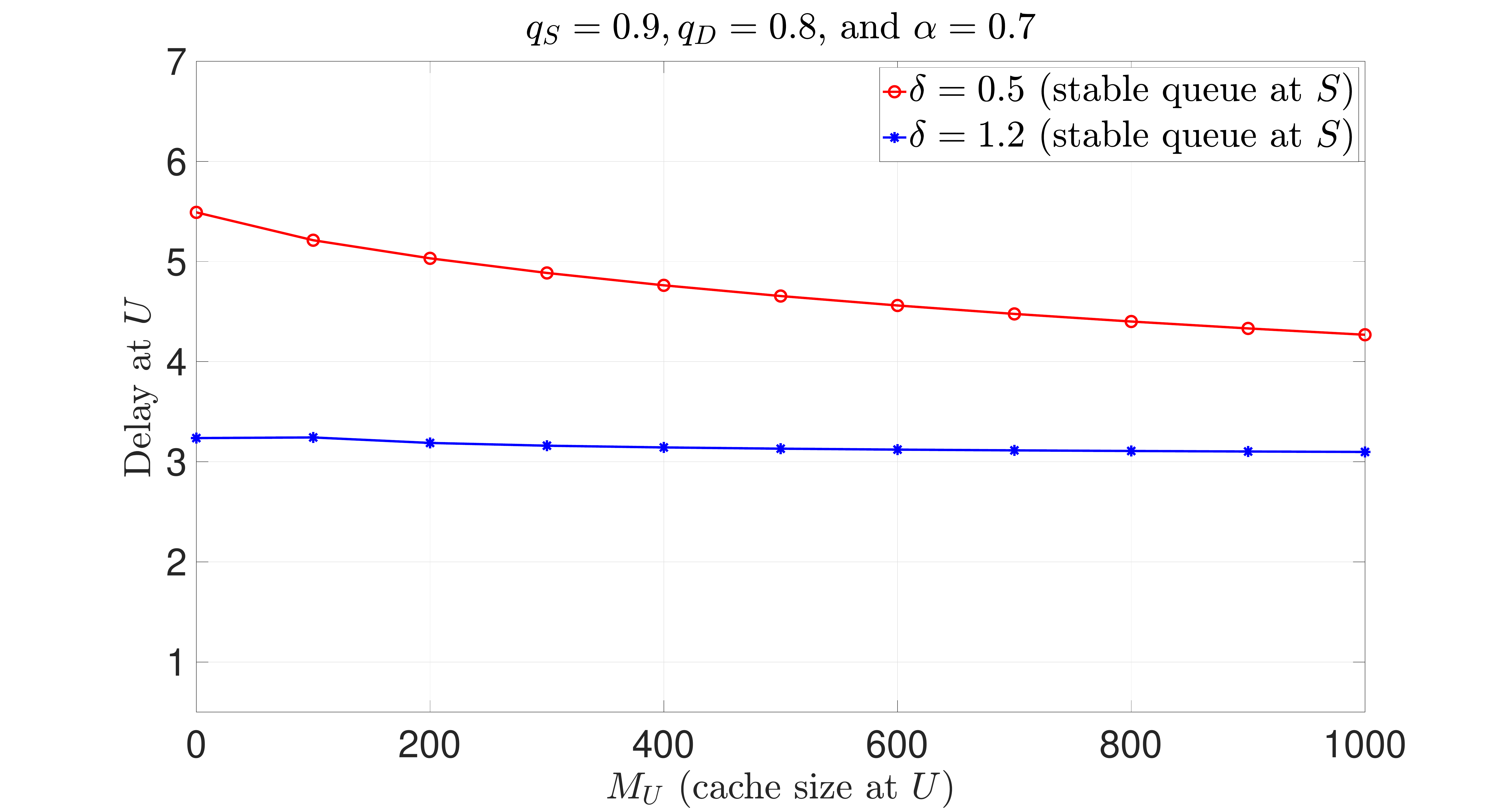}}
	\caption{The average delay at $U$ vs. $M_u$ for $\delta \in \{0.5, 1.2\}$. }
	\label{Fig:delay_vs_Mu}
\end{figure} 

In \figurename~\ref{Fig:delay_vs_Mu}, we show the average delay at $U$ versus the cache size at $U$ for $\lambda=0.2$ and $\lambda=0.4$. 
The cache size $M_U$ affects the request probability for external content, $q_U$, and as it increases, the $q_u$ decreases. 
In any case, the queue at $S$ is stable. 
When $\lambda=0.2$, the effect of $M_U$ on the average delay at $U$ is minor.
However, when average arrival rate $\lambda$ is increased, then increasing cache size at $U$ decreases the average delay especially when $\delta$ is lowered.
\section{Conclusion}\label{Sec:Conclusion}
In this paper, we studied the effect of multiple randomly available caching helpers on a wireless system that serves cachable and non-cachable traffic.
We derived the throughput for a system consisting of a user requesting cachable content from a pair of caching helpers within its proximity or a data center. 
The helpers are assumed to exchange non-cachable content as well as assisting the user's needs for cachable content in a random manner.
We optimized the probabilities by which the helpers assist the user's requests to maximize the system throughput. Moreover, we studied the average delay experienced by the user from the time it requested cachable content till content reception.

Our theoretical and numerical results provide insights concerning the system throughput and the delay behavior of wireless systems serving both cachable and non-cachable content with assistance of multiple randomly available caching helpers.
\bibliographystyle{IEEEtran}
\bibliography{bibliography}{\tiny}

\begin{thebibliography}{10}
\providecommand{\url}[1]{#1}
\csname url@samestyle\endcsname
\providecommand{\newblock}{\relax}
\providecommand{\bibinfo}[2]{#2}
\providecommand{\BIBentrySTDinterwordspacing}{\spaceskip=0pt\relax}
\providecommand{\BIBentryALTinterwordstretchfactor}{4}
\providecommand{\BIBentryALTinterwordspacing}{\spaceskip=\fontdimen2\font plus
\BIBentryALTinterwordstretchfactor\fontdimen3\font minus
  \fontdimen4\font\relax}
\providecommand{\BIBforeignlanguage}[2]{{%
\expandafter\ifx\csname l@#1\endcsname\relax
\typeout{** WARNING: IEEEtran.bst: No hyphenation pattern has been}%
\typeout{** loaded for the language `#1'. Using the pattern for}%
\typeout{** the default language instead.}%
\else
\language=\csname l@#1\endcsname
\fi
#2}}
\providecommand{\BIBdecl}{\relax}
\BIBdecl

\bibitem{Cisco}
\BIBentryALTinterwordspacing
``{Cisco Visual Networking Index: Global Mobile Data Traffic Forecast Update,
  2017–2022 White Paper},'' 2019. [Online]. Available:
  \url{https://www.cisco.com/c/en/us/solutions/collateral/service-provider/visual-networking-index-vni/white-paper-c11-738429.html}
\BIBentrySTDinterwordspacing

\bibitem{Paschos_IEEE_Comm_Mag}
G.~{Paschos}, E.~{Bastug}, I.~{Land}, G.~{Caire}, and M.~{Debbah}, ``Wireless
  caching: technical misconceptions and business barriers,'' \emph{IEEE
  Communications Magazine}, vol.~54, no.~8, pp. 16--22, Aug. 2016.

\bibitem{FemToCaching}
K.~{Shanmugam}, N.~{Golrezaei}, A.~G. {Dimakis}, A.~F. {Molisch}, and
  G.~{Caire}, ``{FemtoCaching: Wireless Content Delivery Through Distributed
  Caching Helpers},'' \emph{IEEE Transactions on Information Theory}, vol.~59,
  no.~12, pp. 8402--8413, Dec. 2013.

\bibitem{Paschos_JSAC}
G.~S. {Paschos}, G.~{Iosifidis}, M.~{Tao}, D.~{Towsley}, and G.~{Caire}, ``{The
  Role of Caching in Future Communication Systems and Networks},'' \emph{IEEE
  Journal on Selected Areas in Communications}, vol.~36, no.~6, pp. 1111--1125,
  June 2018.

\bibitem{Fundamental_Limits_of_Caching}
M.~A. {Maddah-Ali} and U.~{Niesen}, ``{Fundamental Limits of Caching},''
  \emph{IEEE Transactions on Information Theory}, vol.~60, no.~5, pp.
  2856--2867, May 2014.

\bibitem{Prob_Caching_Cache_Hit_vs_Througput_Optimal}
Z.~{Chen}, N.~{Pappas}, and M.~{Kountouris}, ``{Probabilistic Caching in
  Wireless D2D Networks: Cache Hit Optimal Versus Throughput Optimal},''
  \emph{IEEE Communications Letters}, vol.~21, no.~3, pp. 584--587, Mar. 2017.

\bibitem{Delay_Geo_Caching_2tier_HetNets}
E.~{Baştuğ}, M.~{Kountouris}, M.~{Bennis}, and M.~{Debbah}, ``On the delay of
  geographical caching methods in two-tiered heterogeneous networks,''
  \emph{{IEEE 17th International Workshop on Signal Processing Advances in
  Wireless Communications (SPAWC)}}, pp. 1--5, July 2016.

\bibitem{Pappas_Zheng_Access_2019}
N.~{Pappas}, Z.~{Chen}, and I.~{Dimitriou}, ``{Throughput and Delay Analysis of
  Wireless Caching Helper Systems With Random Availability},'' \emph{IEEE
  Access}, vol.~6, pp. 9667--9678, 2018.

\bibitem{Cooperation_Caching_for_HetNets_2017}
J.~{Ma}, J.~{Wang}, and P.~{Fan}, ``{A Cooperation-Based Caching Scheme for
  Heterogeneous Networks},'' \emph{IEEE Access}, vol.~5, pp. 15\,013--15\,020,
  2017.

\bibitem{Cooperative_Caching_Placement_in_Cached_D2D_Underlaid_Cellular_Nets}
Y.~{Wang}, X.~{Tao}, X.~{Zhang}, and Y.~{Gu}, ``{Cooperative Caching Placement
  in Cache-Enabled D2D Underlaid Cellular Network},'' \emph{IEEE Communications
  Letters}, vol.~21, no.~5, pp. 1151--1154, May 2017.

\bibitem{Chen_CooperativeCaching_TWC}
Z.~{Chen}, J.~{Lee}, T.~Q.~S. {Quek}, and M.~{Kountouris}, ``{Cooperative
  Caching and Transmission Design in Cluster-Centric Small Cell Networks},''
  \emph{IEEE Transactions on Wireless Communications}, vol.~16, no.~5, pp.
  3401--3415, May 2017.

\bibitem{FD_communications_2018}
M.~{Naslcheraghi}, M.~{Afshang}, and H.~S. {Dhillon}, ``{Modeling and
  Performance Analysis of Full-Duplex Communications in Cache-Enabled D2D
  Networks},'' \emph{IEEE International Conference on Communications (ICC)},
  pp. 1--6, May 2018.

\bibitem{D2D_vs_Small_Cell_Caching}
Z.~{Chen} and M.~{Kountouris}, ``{D2D caching vs. small cell caching: Where to
  cache content in a wireless network?}'' \emph{IEEE 17th International
  Workshop on Signal Processing Advances in Wireless Communications (SPAWC)},
  pp. 1--6, July 2016.

\bibitem{EE_Downlink_Caching_at_BS}
D.~{Liu} and C.~{Yang}, ``{Energy Efficiency of Downlink Networks With Caching
  at Base Stations},'' \emph{IEEE Journal on Selected Areas in Communications},
  vol.~34, no.~4, pp. 907--922, Apr. 2016.

\bibitem{EE_Wireless_Caching_D2D_Cooperative}
S.~{Lin}, D.~{Cheng}, G.~{Zhao}, and Z.~{Chen}, ``{Energy-Efficient Wireless
  Caching in Device-to-Device Cooperative Networks},'' \emph{IEEE 85th
  Vehicular Technology Conference (VTC Spring)}, June 2017.

\bibitem{Cached_D2D_Communications_Offloading_Gain_and_Energy_Cost}
B.~{Chen}, C.~{Yang}, and A.~F. {Molisch}, ``{Cache-Enabled Device-to-Device
  Communications: Offloading Gain and Energy Cost},'' \emph{IEEE Transactions
  on Wireless Communications}, vol.~16, no.~7, pp. 4519--4536, July 2017.

\bibitem{Optimal_Caching_Placement_for_D2D}
J.~{Rao}, H.~{Feng}, C.~{Yang}, Z.~{Chen}, and B.~{Xia}, ``{Optimal caching
  placement for D2D assisted wireless caching networks},'' \emph{IEEE
  International Conference on Communications (ICC)}, pp. 1--6, May 2016.

\bibitem{Spatially_Caching_for_D2D}
D.~{Malak}, M.~{Al-Shalash}, and J.~G. {Andrews}, ``{Spatially Correlated
  Content Caching for Device-to-Device Communications},'' \emph{{IEEE
  Transactions on Wireless Communications}}, vol.~17, no.~1, pp. 56--70, Jan.
  2018.

\bibitem{EdgeCaching_D2D_offloading}
W.~{Wang}, R.~{Lan}, J.~{Gu}, A.~{Huang}, H.~{Shan}, and Z.~{Zhang}, ``{Edge
  Caching at Base Stations With Device-to-Device Offloading},'' \emph{IEEE
  Access}, vol.~5, pp. 6399--6410, 2017.

\bibitem{Optimal_Caching_and_Scheduling_for_Cache_enabled_D2D_communications}
B.~{Chen}, C.~{Yang}, and Z.~{Xiong}, ``{Optimal Caching and Scheduling for
  Cache-Enabled D2D Communications},'' \emph{IEEE Communications Letters},
  vol.~21, no.~5, pp. 1155--1158, May 2017.

\bibitem{CachingDiversity}
Y.~{Chen} and H.~{Zhang}, ``{Exploiting Transmission and Caching Diversity in
  Cache-Enabled User-Centric Network: Analysis and Optimization},'' \emph{IEEE
  Access}, vol.~7, pp. 65\,934--65\,943, 2019.

\bibitem{OnlineCaching}
K.~{Thar}, N.~H. {Tran}, S.~{Ullah}, T.~Z. {Oo}, and C.~S. {Hong}, ``{Online
  Caching and Cooperative Forwarding in Information Centric Networking},''
  \emph{IEEE Access}, vol.~6, pp. 59\,679--59\,694, 2018.

\bibitem{Single_Bottleneck_Caching_Networks_Analysis}
F.~{Rezaei} and B.~H. {Khalaj}, ``{Stability, Rate, and Delay Analysis of
  Single Bottleneck Caching Networks},'' \emph{IEEE Transactions on
  Communications}, vol.~64, no.~1, pp. 300--313, Jan. 2016.

\bibitem{Adaptive_Video_Steaming_with_Multiple_Helpers}
D.~{Bethanabhotla}, G.~{Caire}, and M.~J. {Neely}, ``{Adaptive Video Streaming
  for Wireless Networks With Multiple Users and Helpers},'' \emph{IEEE
  Transactions on Communications}, vol.~63, no.~1, pp. 268--285, Jan 2015.

\bibitem{VehNets_OnDemandStreaming}
X.~{Hong}, J.~{Jiao}, A.~{Peng}, J.~{Shi}, and C.~{Wang}, ``{Cost Optimization
  for On-Demand Content Streaming in IoV Networks With Two Service Tiers},''
  \emph{{IEEE Internet of Things Journal}}, vol.~6, no.~1, pp. 38--49, Feb.
  2019.

\bibitem{VehNets_qlearning}
L.~{Hou}, L.~{Lei}, K.~{Zheng}, and X.~{Wang}, ``{A Q-Learning based Proactive
  Caching Strategy for Non-safety Related Services in Vehicular Networks},''
  \emph{IEEE Internet of Things Journal}, pp. 1--1, 2019.

\bibitem{MEC_survey}
S.~{Wang}, X.~{Zhang}, Y.~{Zhang}, L.~{Wang}, J.~{Yang}, and W.~{Wang}, ``{A
  Survey on Mobile Edge Networks: Convergence of Computing, Caching and
  Communications},'' \emph{IEEE Access}, vol.~5, pp. 6757--6779, 2017.

\bibitem{CachingScheme_MEC}
X.~{Liu}, J.~{Zhang}, X.~{Zhang}, and W.~{Wang}, ``{Mobility-Aware Coded
  Probabilistic Caching Scheme for MEC-Enabled Small Cell Networks},''
  \emph{IEEE Access}, vol.~5, pp. 17\,824--17\,833, 2017.

\bibitem{CachingChallenges}
Y.~{Wu}, S.~{Yao}, Y.~{Yang}, T.~{Zhou}, H.~{Qian}, H.~{Hu}, and
  M.~{Hamalainen}, ``{Challenges of Mobile Social Device Caching},'' \emph{IEEE
  Access}, vol.~4, pp. 8938--8947, 2016.

\bibitem{VehNets_LowLatencyRadioAccess}
S.~{Lien}, S.~{Hung}, D.~{Deng}, C.~{Lai}, and H.~{Tsai}, ``{Low Latency Radio
  Access in 3GPP Local Area Data Networks for V2X: Stochastic Optimization and
  Learning},'' \emph{{IEEE Internet of Things Journal}}, pp. 1--1, 2019.

\bibitem{Advances_Edge_IoT}
Z.~{Piao}, M.~{Peng}, Y.~{Liu}, and M.~{Daneshmand}, ``{Recent Advances of Edge
  Cache in Radio Access Networks for Internet of Things: Techniques,
  Performances, and Challenges},'' \emph{{IEEE Internet of Things Journal}},
  vol.~6, no.~1, pp. 1010--1028, Feb 2019.

\bibitem{JointOptimization1}
S.~{Sardellitti}, G.~{Scutari}, and S.~{Barbarossa}, ``{Joint Optimization of
  Radio and Computational Resources for Multicell Mobile-Edge Computing},''
  \emph{{IEEE Transactions on Signal and Information Processing over
  Networks}}, vol.~1, no.~2, pp. 89--103, June 2015.

\bibitem{JointOptimization2}
S.~{Barbarossa}, S.~{Sardellitti}, and P.~{Di Lorenzo}, ``{Joint allocation of
  computation and communication resources in multiuser mobile cloud
  computing},'' in \emph{{IEEE 14th Workshop on Signal Processing Advances in
  Wireless Communications (SPAWC)}}, June 2013, pp. 26--30.

\bibitem{Fog_JointCachingComputing}
Y.~{Wei}, F.~R. {Yu}, M.~{Song}, and Z.~{Han}, ``{Joint Optimization of
  Caching, Computing, and Radio Resources for Fog-Enabled IoT using Natural
  Actor–Critic Deep Reinforcement Learning},'' \emph{{IEEE Internet of Things
  Journal}}, vol.~6, no.~2, pp. 2061--2073, Apr. 2019.

\bibitem{CRAN_ContentPlacement}
J.~{Yao} and N.~{Ansari}, ``{Joint Content Placement and Storage Allocation in
  C-RANs for IoT Sensing Service},'' \emph{{IEEE Internet of Things Journal}},
  vol.~6, no.~1, pp. 1060--1067, Feb 2019.

\bibitem{SmartGrid_caching}
X.~{Huang} and N.~{Ansari}, ``{Content Caching and Distribution in Smart Grid
  Enabled Wireless Networks},'' \emph{{IEEE Internet of Things Journal}},
  vol.~4, no.~2, pp. 513--520, Apr. 2017.

\bibitem{CaaS}
X.~{Li}, X.~{Wang}, K.~{Li}, and V.~C.~M. {Leung}, ``{CaaS: Caching as a
  Service for 5G Networks},'' \emph{IEEE Access}, vol.~5, pp. 5982--5993, 2017.

\bibitem{Pappas_FD_TWC}
N.~{Pappas}, M.~{Kountouris}, A.~{Ephremides}, and A.~{Traganitis},
  ``{Relay-Assisted Multiple Access With Full-Duplex Multi-Packet Reception},''
  \emph{IEEE Transactions on Wireless Communications}, vol.~14, no.~7, pp.
  3544--3558, July 2015.

\bibitem{loynes_1962}
R.~M. Loynes, ``The stability of a queue with non-independent inter-arrival and
  service times,'' \emph{Mathematical Proceedings of the Cambridge
  Philosophical Society}, vol.~58, no.~3, pp. 497--520, 1962.

\bibitem{Walrand}
J.~Walrand, \emph{{Communication Networks: A First Course}}.\hskip 1em plus
  0.5em minus 0.4em\relax {2nd ed., New York, NY, USA,: McGraw-Hill}, 1998.

\end{thebibliography}

\end{document}